\documentclass[12pt]{article}
\usepackage{latexsym,amsmath,amsfonts,amssymb}
\usepackage[latin1]{inputenc}
\usepackage[american]{babel}
\usepackage[dvips]{graphicx}
\usepackage{bbm}
\newcommand{\be}{\begin{equation}}
\newcommand{\ee}{\end{equation}}
\newcommand{\beq}{\begin{equation}}
\newcommand{\eeq}{\end{equation}}
\newcommand{\bea}{\begin{eqnarray}}
\newcommand{\eea}{\end{eqnarray}}

\newcommand{\ba}{\begin{eqnarray}}
\newcommand{\ea}{\end{eqnarray}}
\begin{document}
\baselineskip=15.5pt
\pagestyle{plain}
\setcounter{page}{1}


\def\del{{\partial}}
\def\vev#1{\left\langle #1 \right\rangle}
\def\cn{{\cal N}}
\def\co{{\cal O}}
\def\IC{{\mathbb C}}
\def\IR{{\mathbb R}}
\def\IZ{{\mathbb Z}}
\def\RP{{\bf RP}}
\def\CP{{\bf CP}}
\def\Poincare{{Poincar\'e }}
\def\tr{{\rm tr}}
\def\tp{{\tilde \Phi}}

\def\TL{\hfil$\displaystyle{##}$}
\def\TR{$\displaystyle{{}##}$\hfil}
\def\TC{\hfil$\displaystyle{##}$\hfil}
\def\TT{\hbox{##}}
\def\HLINE{\noalign{\vskip1\jot}\hline\noalign{\vskip1\jot}}
\def\seqalign#1#2{\vcenter{\openup1\jot
   \halign{\strut #1\cr #2 \cr}}}
\def\lbldef#1#2{\expandafter\gdef\csname #1\endcsname {#2}}
\def\eqn#1#2{\lbldef{#1}{(\ref{#1})}%
\begin{equation} #2 \label{#1} \end{equation}}
\def\eqalign#1{\vcenter{\openup1\jot
     \halign{\strut\span\TL & \span\TR\cr #1 \cr
    }}}
\def\eno#1{(\ref{#1})}
\def\href#1#2{#2}
\def\half{{1 \over 2}}

\def\ads{{\it AdS}}
\def\adsp{{\it AdS}$_{p+2}$}
\def\cft{{\it CFT}}

\newcommand{\ber}{\begin{eqnarray}}
\newcommand{\eer}{\end{eqnarray}}

\newcommand{\beqar}{\begin{eqnarray}}
\newcommand{\cN}{{\cal N}}
\newcommand{\cO}{{\cal O}}
\newcommand{\cA}{{\cal A}}
\newcommand{\cT}{{\cal T}}
\newcommand{\cF}{{\cal F}}
\newcommand{\cC}{{\cal C}}
\newcommand{\cR}{{\cal R}}
\newcommand{\cW}{{\cal W}}
\newcommand{\eeqar}{\end{eqnarray}}
\newcommand{\tht}{\thteta}
\newcommand{\lm}{\lambda}\newcommand{\Lm}{\Lambda}
\newcommand{\eps}{\epsilon}


\newcommand{\nonu}{\nonumber}
\newcommand{\oh}{\displaystyle{\frac{1}{2}}}
\newcommand{\dsl}
   {\kern.06em\hbox{\raise.15ex\hbox{$/$}\kern-.56em\hbox{$\partial$}}}
\newcommand{\id}{i\!\!\not\!\partial}
\newcommand{\as}{\not\!\! A}
\newcommand{\ps}{\not\! p}
\newcommand{\ks}{\not\! k}
\newcommand{\D}{{\cal{D}}}
\newcommand{\dv}{d^2x}
\newcommand{\Z}{{\cal Z}}
\newcommand{\N}{{\cal N}}
\newcommand{\Dsl}{\not\!\! D}
\newcommand{\Bsl}{\not\!\! B}
\newcommand{\Psl}{\not\!\! P}
\newcommand{\eeqarr}{\end{eqnarray}}
\newcommand{\ZZ}{{\rm \kern 0.275em Z \kern -0.92em Z}\;}


\def\del{{\delta^{\hbox{\sevenrm B}}}} \def\ex{{\hbox{\rm e}}}
\def\azb{A_{\bar z}} \def\az{A_z} \def\bzb{B_{\bar z}} \def\bz{B_z}
\def\czb{C_{\bar z}} \def\cz{C_z} \def\dzb{D_{\bar z}} \def\dz{D_z}
\def\im{{\hbox{\rm Im}}} \def\mod{{\hbox{\rm mod}}} \def\tr{{\hbox{\rm Tr}}}
\def\ch{{\hbox{\rm ch}}} \def\imp{{\hbox{\sevenrm Im}}}
\def\trp{{\hbox{\sevenrm Tr}}} \def\vol{{\hbox{\rm Vol}}}
\def\rl{\Lambda_{\hbox{\sevenrm R}}} \def\wl{\Lambda_{\hbox{\sevenrm W}}}
\def\fc{{\cal F}_{k+\cox}} \def\vev{vacuum expectation value}
\def\nodiv{\mid{\hbox{\hskip-7.8pt/}}}
\def\ie{{\em i.e.}}
\def\ie{\hbox{\it i.e.}}

\def\CC{{\mathchoice
{\rm C\mkern-8mu\vrule height1.45ex depth-.05ex
width.05em\mkern9mu\kern-.05em}
{\rm C\mkern-8mu\vrule height1.45ex depth-.05ex
width.05em\mkern9mu\kern-.05em}
{\rm C\mkern-8mu\vrule height1ex depth-.07ex
width.035em\mkern9mu\kern-.035em}
{\rm C\mkern-8mu\vrule height.65ex depth-.1ex
width.025em\mkern8mu\kern-.025em}}}

\def\RR{{\rm I\kern-1.6pt {\rm R}}}
\def\NN{{\rm I\!N}}
\def\ZZ{{\rm Z}\kern-3.8pt {\rm Z} \kern2pt}
\def\IB{\relax{\rm I\kern-.18em B}}
\def\ID{\relax{\rm I\kern-.18em D}}
\def\II{\relax{\rm I\kern-.18em I}}
\def\IP{\relax{\rm I\kern-.18em P}}
\newcommand{\CS}{{\scriptstyle {\rm CS}}}
\newcommand{\CSs}{{\scriptscriptstyle {\rm CS}}}
\newcommand{\rc}{\nonumber\\}
\newcommand{\bear}{\begin{eqnarray}}
\newcommand{\eear}{\end{eqnarray}}
\newcommand{\W}{{\cal W}}
\newcommand{\F}{{\cal F}}
\newcommand{\x}{{\cal O}}
\newcommand{\LL}{{\cal L}}

\def\mani{{\cal M}}
\def\calo{{\cal O}}
\def\calb{{\cal B}}
\def\calw{{\cal W}}
\def\calz{{\cal Z}}
\def\cald{{\cal D}}
\def\calc{{\cal C}}
\def\to{\rightarrow}
\def\ele{{\hbox{\sevenrm L}}}
\def\ere{{\hbox{\sevenrm R}}}
\def\zb{{\bar z}}
\def\wb{{\bar w}}
\def\nodiv{\mid{\hbox{\hskip-7.8pt/}}}
\def\menos{\hbox{\hskip-2.9pt}}
\def\dr{\dot R_}
\def\drr{\dot r_}
\def\ds{\dot s_}
\def\da{\dot A_}
\def\dga{\dot \gamma_}
\def\ga{\gamma_}
\def\dal{\dot\alpha_}
\def\al{\alpha_}
\def\cl{{closed}}
\def\cls{{closing}}
\def\vev{vacuum expectation value}
\def\tr{{\rm Tr}}
\def\to{\rightarrow}
\def\too{\longrightarrow}


\def\a{\alpha}
\def\b{\beta}
\def\c{\gamma}
\def\d{\delta}
\def\e{\epsilon}           
\def\f{\phi}               
\def\vf{\varphi}  \def\tvf{\tilde{\varphi}}
\def\vp{\varphi}
\def\g{\gamma}
\def\h{\eta}
\def\i{\iota}
\def\j{\psi}
\def\k{\kappa}                    
\def\l{\lambda}
\def\m{\mu}
\def\n{\nu}
\def\o{\omega}  \def\w{\omega}
\def\q{\theta}  \def\th{\theta}                  
\def\r{\rho}                                     
\def\s{\sigma}                                   
\def\t{\tau}
\def\u{\upsilon}
\def\x{\xi}
\def\z{\zeta}
\def\pt{\tilde{\varphi}}
\def\tt{\tilde{\theta}}
\def\lab{\label}
\def\6{\partial}
\def\wg{\wedge}
\def\atanh{{\rm arctanh}}
\def\bpsi{\bar{\psi}}
\def\bt{\bar{\theta}}
\def\bvf{\bar{\varphi}}

%

\newfont{\namefont}{cmr10}
\newfont{\addfont}{cmti7 scaled 1440}
\newfont{\boldmathfont}{cmbx10}
\newfont{\headfontb}{cmbx10 scaled 1728}
\newcommand{\re}{\,\mathbb{R}\mbox{e}\,}
\newcommand{\hyph}[1]{$#1$\nobreakdash-\hspace{0pt}}
\providecommand{\abs}[1]{\lvert#1\rvert}
\newcommand{\Nugual}[1]{$\mathcal{N}= #1 $}
\newcommand{\sub}[2]{#1_\text{#2}}
\newcommand{\partfrac}[2]{\frac{\partial #1}{\partial #2}}
\newcommand{\bsp}[1]{\begin{equation} \begin{split} #1 \end{split} \end{equation}}
\newcommand{\calF}{\mathcal{F}}
\newcommand{\calO}{\mathcal{O}}
\newcommand{\calM}{\mathcal{M}}
\newcommand{\calV}{\mathcal{V}}
\newcommand{\bbZ}{\mathbb{Z}}
\newcommand{\bbC}{\mathbb{C}}

\numberwithin{equation}{section}

\newcommand{\Tr}{\mbox{Tr}}    


%
\renewcommand{\theequation}{{\rm\thesection.\arabic{equation}}}
\begin{titlepage}
\rightline{SISSA 36/2007/EP}
\rightline{US-FT-3/07}
\vspace{0.1in}

\begin{center}
\Large \bf Backreacting Flavors in the Klebanov-Strassler Background
\end{center}
\vskip 0.2truein
\begin{center}
Francesco Benini${}^{*}$\footnote{benini@sissa.it},
Felipe Canoura$^{**}$\footnote{canoura@fpaxp1.usc.es}, Stefano
Cremonesi${}^{*}$\footnote{cremones@sissa.it},
Carlos
N\'u\~nez${}^{\dagger}$\footnote{c.nunez@swansea.ac.uk} \\ and
Alfonso V. Ramallo ${}^{**}$\footnote{alfonso@fpaxp1.usc.es}
\vspace{0.3in}\\
${}^{*}$ \it{ SISSA/ISAS and INFN-Sezione di Trieste\\ Via Beirut 2;
I-34014 Trieste, Italy}
\vspace{0.2in}
\vskip 0.1truein
${}^{**}$
\it{
Departamento de  Fisica de Particulas, Universidade
de Santiago de
Compostela\\and\\Instituto Galego de Fisica de Altas
Enerxias (IGFAE)\\E-15782, Santiago de Compostela, Spain
}
\vspace{0.2in}
\vskip 0.1truein
${}^{\dagger}$ \it{Department of Physics\\ University of Swansea, Singleton
Park\\
Swansea SA2 8PP, United Kingdom.}
\vspace{0.2in}
\end{center}
\vspace{0.2in}
\centerline{{\bf Abstract}}
In this paper we present new analytic solutions of pure Type IIB supergravity plus D7-branes describing the addition of an arbitrary number of flavors to the Klebanov-Tseytlin and Klebanov-Strassler backgrounds. We provide a precise field theory dual and a detailed analysis of the duality cascade which describes its RG flow. Matchings of beta functions and anomalies between the field theory and the string setup are presented. 
We give an understanding of Seiberg duality as a large gauge transformation on the RR and NSNS potentials. We also analyze the UV behavior of the field theory, that as suggested by the string background can be associated with a duality wall.

\smallskip
\end{titlepage}
\setcounter{footnote}{0}

\newpage

\section{Introduction}
Early ideas of 't Hooft \cite{'tHooft:1973jz} and the experimental evidence for stringy behavior in hadronic physics 
suggested that aspects of the strong interaction can be described, 
predicted, and understood using a (not yet known) string theory. These 
ideas started to materialize when the Maldacena conjecture (also known as 
AdS/CFT correspondence) and some of its refinements were formulated 
\cite{Maldacena:1997re,Itzhaki:1998dd,Gubser:1998bc,Witten:1998qj}. 
Indeed, a (3+1)-dimensional field theory was 
shown to contain strings that captured non-perturbative and perturbative 
physics. The downside was that the field theory in question ($\mathcal{N}=4$ SYM) 
was not of immediate relevance to hadronic physics. The 
necessity of finding extensions of these ideas to phenomenologically
more interesting field theories was then well motivated.

A very fruitful extension of the AdS/CFT correspondence stems from 
studying branes at conical singularities, of which the case of D3-branes 
on conifolds is a particular example presenting especially rich 
dynamics. The study of this system was developed in a list of many 
interesting papers: those that have been more influential to the 
present work are \cite{Klebanov:1998hh,Gubser:1998fp,Klebanov:1999rd,Klebanov:2000nc,Klebanov:2000hb,Gubser:2004qj,Dymarsky:2005xt,Benna:2006ib,Butti:2004pk,Strassler:2005qs}. 

In this paper, we will consider the addition of new degrees of freedom to 
the Klebanov-Tseytlin(KT) \cite{Klebanov:2000nc} and Klebanov-Strassler (KS)
 \cite{Klebanov:2000hb} solutions. 
These new excitations will be incorporated in the form of D7 flavor 
branes, corresponding to fundamental matter in the dual field theory. Unlike the color branes, which disappear in the geometric transition and are substituted by closed string fluxes, the flavor branes appear explicitly in our solutions: they correspond to the open strings which are suggested by Veneziano's topological expansion of large N gauge theories \cite{Veneziano:1976wm}.   
The addition of flavors to these field theories was first considered in \cite{Karch:2002sh,Ouyang:2003df,Kuperstein:2004hy}. 
We will follow ideas introduced in \cite{Karch:2002sh}, but will consider the case in which the number of 
fundamental fields is of the same order as the number of 
adjoint or bifundamental fields, that is $N_f\sim N_c$. This means that the new (strongly 
coupled) dynamics of the field theory is captured by a background that includes the 
backreaction of the flavor branes. In order to find the new solutions, we 
follow the ideas and techniques of \cite{Casero:2006pt,Paredes:2006wb,Casero:2007pz,Benini:2006hh}.

Let us describe the main achievements of this paper.
We will present {\it analytic} solutions for the equations of motion of Type IIB 
supergravity coupled to the DBI+WZ action of the flavor D7-branes that preserve 
minimal SUSY in four dimensions; we show how to reduce these solutions to those found by Klebanov-Tseytlin/Strassler when the number of flavors is taken to 
zero. Using them, we make a precise matching between the field theory cascade (that, enriched by the presence of the fundamentals, is still self-similar) and the string predictions. We 
will also match anomalies and beta functions by using our 
new supergravity background. There is a  variety of other 
field theory observables that can be predicted by our solutions, and
will be presented in a future publication  \cite{us}, where also many technical details 
and nice subtleties of the present paper will be explained.

The organization of this paper is as follows. 
In Section \ref{section-ansatz} we present the setup, the ansatz and the strategy to find supersymmetric solutions of the Bianchi identities and equations of motion. We also introduce the notion of the so-called Page charges and we analyze their values in our particular ansatz. In Sections \ref{KSflavored} and \ref{KTflavored} we present two main solutions, which reflect the addition of flavors to the KS and KT 
backgrounds, respectively. 
In Section \ref{sect: Field Th} we present the dual field theory and propose that its RG flow can be understood in terms of a cascade of Seiberg dualities.
In Section \ref{cascade-SUGRA} we show that the duality cascade is encoded in our supergravity solutions, by comparing ranks of the groups in field theory with effective charges in supergravity, and matching R-anomalies and beta functions of gauge couplings on both sides of the gauge/gravity duality.
The behavior of the background in the UV of the gauge theory suggests that the field theory generates a `duality wall'. We also give a nice picture of Seiberg duality as a large gauge transformation in supergravity.
We close the paper with possible future directions in Section \ref{Sect: final remarks}.

\section{The setup and the ansatz}\label{section-ansatz}
We are interested in adding to the KT/KS cascading gauge theory \cite{Klebanov:2000nc,Klebanov:2000hb} a number of flavors (fundamental fields) comparable with the number of colors (adjoint and bifundamental fields). Those supergravity backgrounds were obtained by considering a stack of regular and fractional D3-branes at the tip of the conifold. After the geometric transition, the color branes disappear from the geometry, but the closed string fluxes they sourced remain nontrivial. The addition of flavors in the field theory, in the large N limit considered by Veneziano \cite{Veneziano:1976wm}, amounts to introduce mesonic currents and internal quark loops in the planar diagrams that survive 't Hooft's double scaling limit \cite{'tHooft:1973jz}: Feynman diagrams in Veneziano's limit arrange in a topological expansion reminiscent of the loop expansion of an open and closed string theory, where the new open string sector arises from loops of flavor fields in the field theory. 
The dual string picture of Veneziano's large N limit involves the addition of backreacting `flavor' branes to the backgrounds which describe the 't Hooft limit of some gauge theories.

Let us then consider a system of type IIB supergravity plus $N_f$ D7-branes.
The dynamics of the latter will be governed by the corresponding
Dirac-Born-Infeld and Wess-Zumino actions. Our solution will have a non-trivial metric and
dilaton $\phi$ and, as in any cascading background,  non-vanishing
RR three- and five-forms $F_3$ and $F_5$, as well as a non-trivial NSNS
three-form $H_3$. In addition, the D7-branes act as a source for (the Hodge
dual of) the RR one-form $F_1$ through the WZ coupling:
\begin{equation}
S_{WZ}^{D7}\,=\,T_7\,\sum_{N_f}\,\int_{{\cal
M}_8}\,\hat C_{8}\,+\,\cdots\,\,,
\end{equation}
which generically induces a violation of the Bianchi identity $dF_1=0$.
Therefore our configuration will also  necessarily have a
non-vanishing value of $F_1$. The ansatz we shall adopt for the
Einstein frame metric is the following:
\bear
\label{metric}
&&ds^2\,=\,\Big[\,h(r)\,\Big]^{-\frac{1}{2}}\,dx^2_{1,3}\,+
\,\Big[\,h(r)\,\Big]^{\frac{1}{2}}\,\Bigg[\,dr^2\,+\,
e^{2G_1(r)}\,(\sigma_1^2\,+\,\sigma_2^2)\,+\,\,
\rc
&&\,+\,e^{2G_2(r)} \bigg(
(\omega_1\,+\,g(r)\,\sigma_1)^2\,+\,(\omega_2\,+\,g(r)\,\sigma_2)^2
\bigg)\,+\,{{e^{2G_3(r)}}\over 9}\,
\big(\omega_3\,+\,\sigma_3)^2\,\Bigg ] \,\, ,
\eear
where $dx^2_{1,3}$ denotes the four-dimensional Minkowski metric and
$\sigma_i$ and $\omega_i$ ($i=1,2,3$) are one-forms that can be
written in terms of  five angular coordinates
$(\theta_1, \varphi_1, \theta_2, \varphi_2,\psi)$ as follows:
\bear
&& \sigma_1\,=\,d\theta_1 \, , \qquad \qquad
\sigma_2\,=\,\sin{\theta_1}\,
d\varphi_1 \, , \qquad \qquad
\sigma_3\,=\,\cos{\theta_1}\,d\varphi_1 \, , \rc
&&\omega_1\,=\,\sin{\psi} \sin{\theta_2}\,
d\varphi_2\,+\,\cos{\psi}\,d\theta_2\,\, ,
\qquad\qquad
\omega_2\,=\,-\cos{\psi}
\sin{\theta_2}\, d\varphi_2\,+\,\sin{\psi}\,d\theta_2\,\, , \rc
&&\omega_3\,=\,d\psi\,+\,\cos{\theta_2}\,d\varphi_2 \,\, .
\eear
Notice that our metric ansatz  (\ref{metric}) depends on five unknown
radial functions $G_i(r)$ ($i=1,2,3$), $g(r)$ and $h(r)$. The ansatz for
$F_5$ has the standard form, namely:
\begin{equation}
\label{F5}
F_5\,=\,d h^{-1}(r)\wedge dx^0\wedge\cdots\wedge dx^3\,+\,{\rm Hodge\,\,dual} \;.
\end{equation}
As expected for flavor branes, we will take D7-branes extended along the
four Minkowski coordinates as well as other four internal coordinates. The $\kappa$-symmetric embedding of the D7-branes we start from will be discussed in Section \ref{sect: Field Th}. 
In order to simplify the computations, following the approach of \cite{Benini:2006hh}, we will smear
the D7-branes in their two transverse directions in such a way that the symmetries of the unflavored background are recovered. As explained in \cite{Benini:2006hh}, this smearing amounts to the following
generalization of the WZ term of the D7-brane action:
\begin{equation}
\label{WZ-smearing}
S_{WZ}^{D7}\,=\,T_7\,\,\sum_{N_f}\,\,\int_{{\cal M}_8}\,\,\hat C_8\,\,+\,
\cdots
\qquad \rightarrow \qquad
\,\,T_7\,\,\int_{{\cal M}_{10}}\,
\Omega\wedge C_8\,\,+\,
\cdots,
\end{equation}
where $\Omega$ is a two-form which determines the distribution of the RR
charge of the D7-brane and ${\cal M}_{10}$ is the full ten-dimensional
manifold. Notice that $\Omega$ acts as a magnetic charge source for $F_1$
which generates  a violation of its Bianchi identity. Actually, from the
equation of motion of $C_8$ one gets:
\begin{equation} \label{Bianchi-F1}
dF_1\,= -\Omega\,\,.
\end{equation}
In what follows we will assume that the flavors introduced by the
D7-brane are massless, which is equivalent to require that the flavor
brane worldvolume reaches the origin in the holographic direction. Under
this condition one expects a radial coordinate independent D7-brane charge density. 
Moreover, the D7-brane  embeddings that we will smear imply that $\Omega$ is symmetric
under the exchange of the two $S^2$'s parameterized  by
$(\theta_1, \varphi_1)$ and $(\theta_2, \varphi_2)$, and independent of
$\psi$ (see Section \ref{sect: Field Th}). The smeared charge density distribution is the one already
adopted in \cite{Benini:2006hh}, namely:
\begin{equation} \label{smearing}
dF_1\,= - {N_f\over 4\pi}\,(\sin\theta_1\,d\theta_1\wedge d\varphi_1\,+\,
\sin\theta_2\,d\theta_2\wedge d\varphi_2\,)\,=\,
{N_f\over 4\pi}\,\, (\omega_1 \wedge \omega_2 \,-\,
\sigma_1 \wedge \sigma_2)\,\,,
\end{equation}
where the coefficient $N_f/4\pi$ is determined by normalization. With
this ansatz for $\Omega$ the modified Bianchi identity
(\ref{Bianchi-F1}) determines the value of $F_1$, namely:
\begin{equation}  \label{F1}
F_1\,=\,{N_f\over 4\pi}(\omega_3\,+\,\sigma_3)\,\,.
\end{equation}

The ansatz for the RR and NSNS 3-forms that we propose is an extension of the one
given by Klebanov and Strassler and it is simply (in this paper we set for convenience $\alpha'=1$):
\begin{equation}\begin{aligned}  \label{ansatz}
B_2 &= \frac{M}{2} \Bigl[ f\, g^1 \wedge g^2\,+\,k\, g^3 \wedge 
g^4 \Bigr]  \,\,, \\
H_3 &= dB_2\,=\, \frac{M}{2} \, \Bigl[ dr \wedge (f' \,g^1 \wedge g^2\,+\,
k'\,g^3 \wedge g^4)\,+\,{1 \over 2}(k-f)\, g^5 \wedge (g^1
\wedge g^3\,+\,g^2 \wedge g^4) \Bigr]\,\,,   \\
F_3 &= \frac{M}{2} \Big\{ g^5\wedge \Big[ \big( F+\frac{N_f}{4\pi}f\big)g^1\wedge g^2 + \big(1- F+\frac{N_f}{4\pi}k\big)g^3\wedge g^4 \Big] +F' dr \wedge \big(g^1\wedge g^3 + g^2\wedge g^4   \big)\Big\} , 
\end{aligned} \end{equation}
where $M$ is a constant, $f(r)$, $k(r)$ and $F(r)$ are functions of the radial coordinate, and the  $g^i$'s are the set of one-forms
\bear
&&g^1\,=\,{1 \over {\sqrt{2}}} (\omega_2 \,-	\,\sigma_2)\,\, ,
\qquad \qquad g^2\,=\,{1 \over {\sqrt{2}}} (-\omega_1 \,+\,\sigma_1)\,\, \,\, ,
\rc\rc &&g^3\,=\,{-1 \over {\sqrt{2}}} (\omega_2 \,+\,\sigma_2)\,\, , \qquad
\qquad g^4\,=\,{1 \over {\sqrt{2}}} (\omega_1 \,+\,\sigma_1)\,\, , \rc\rc
&&g^5\,=\, \omega_3 \,+\,\sigma_3\,\, .
\eear
The forms $F_3$, $H_3$ and $F_5$ must satisfy the following set of Bianchi
identities:
\begin{equation}
dF_3\,=\,H_3 \wedge F_1 \,\, ,
\qquad\qquad
dH_3\,=\,0 \,\, ,
\qquad\qquad
dF_5\,=\, \, H_3 \wedge F_3 \,\, .
\end{equation}
Notice that the equations for $F_3$ and $H_3$ are automatically satisfied
by our ansatz (\ref{ansatz}). However, the Bianchi identity for $F_5$
gives rise to the following differential equation:
\begin{equation}
{d \over {dr}} \Big [  h'\, e^{2G_1+2G_2+G_3} \Big ] = -
{3 \over 4}M^2
\Big[ (1-F\,+\,{N_f\over 4\pi}\,k)f'+(F+{N_f\over 4\pi}f)k'
+(k-f) F'\Big ]\,\, ,
\end{equation}
which can be integrated, with the result:
\begin{equation}
 h'\, e^{2G_1+2G_2+G_3}\,=\,-
{3 \over 4}M^2\Big[f-(f-k)F+{N_f\over 4\pi}fk\Big]\,+\,
{\rm constant}\,\,.
\label{Kh2-fkF}
\end{equation}
Let us now parameterize $F_5$ as
\begin{equation}
F_5\,=\,{\pi\over 4}\,N_{eff} (r)\,
g^1\wedge g^2\wedge g^3\wedge g^4\wedge g^5\,+\,
{\rm Hodge\,\,dual}\,\,,
\label{F5-Neff}
\end{equation}
and let us define the five-manifold ${\cal M}_5$ as the one that is obtained
by taking the Minkowski coordinates and $r$ fixed to a constant value.
As $\int_{{\cal M}_5}F_5\,=\,(4\pi^2)^2\,N_{eff} (r)$, it follows that
$N_{eff} (r)$ can be interpreted as the effective D3-brane charge at the
value $r$ of the holographic coordinate. From our ansatz (\ref{F5}), it
follows that:
\begin{equation}
N_{eff} (r)\,=   \,-{4\over 3\pi}\, h'\, e^{2G_1+2G_2+G_3}\,\,,
\label{Neff-Kh2}
\end{equation}
and taking into account (\ref{Kh2-fkF}), we can write
\begin{equation} \label{Neff}
N_{eff} (r) \equiv \frac{1}{(4\pi^2)^2} \int_{{\cal M}_5} F_5 \,=  \,N_0\,+\,{M^2\over \pi}\,\Big[\,
f-(f-k)F+{N_f\over 4\pi}fk\Big]\,\,,
\end{equation}
where $N_0$ is a constant. It follows from \eqref{Neff} that the RR
five-form $F_5$ is determined once the functions $F$, $f$ and $k$ that
parameterize the three-forms are known. Moreover, eq. \eqref{Kh2-fkF} allows to compute the warp factor once the functions $G_i$ and the three-forms are determined. Notice also that the effective D5-brane charge is obtained by integrating the gauge-invariant field strength $F_3$ over 
the 3-cycle $S^3$: $\theta_2=\text{const.}$, $\varphi_2=\text{const.}$. The result is:
\begin{equation} \label{Meff}
M_{eff}(r) \equiv \frac{1}{4\pi^2} \int_{S^3} F_3 = M \Bigl[ 1 + \frac{N_f}{4\pi} (f+k) \Bigr]\;.
\end{equation}

The strategy to proceed further
is to look at  the conditions imposed by supersymmetry. We will smear,
as in \cite{Benini:2006hh}, D7-brane embeddings that are $\kappa$-symmetric and,
therefore, the supersymmetry requirement is equivalent to the
vanishing of the variations of the dilatino and gravitino of type IIB
supergravity under  supersymmetry transformations. These conditions give
 rise to a large number of BPS first order ordinary differential equations for
the dilaton and the different functions that parameterize the metric and
the forms. 
In the end, one can check that the first order differential equations imposed by supersymmetry imply the second order differential equations of motion.
In particular, from the variation of the dilatino we get the
following differential equation for the dilaton:
\begin{equation}
\phi'\,=\,{3N_f\over 4\pi}\,e^{\phi-G_3}\;.
\end{equation}
A detailed analysis of the conditions imposed by supersymmetry shows that
the fibering function $g$ in eq. \eqref{metric} is subjected to the following algebraic
constraint:
\begin{equation}
g \Big [ g^2\,-\,1\,+\,e^{2(G_1-G_2)} \Big ] \,=\,0 \,\,,
\label{g-constraint}
\end{equation}
which has obviously two solutions. The first of these solutions is $g=0$
and, as it is clear from our metric ansatz (\ref{metric}), it corresponds
to the cases of the flavored singular and resolved conifolds. 
 In the second solution $g$ is such
that the term in brackets on the right-hand side of (\ref{g-constraint})
vanishes. This solution gives rise to the flavored version of the warped
deformed conifold. 
The flavored KT solution will be presented in Section \ref{KTflavored}, whereas the flavored KS solution will be analyzed in Section \ref{KSflavored}.

\subsection{Maxwell and Page charges}
\medskip

Before presenting the explicit solutions for the metric and the forms of the supergravity equations, let us discuss the different charges carried out by our solutions. In theories, like type IIB supergravity, that have  Chern-Simons terms in the action (which give rise to modified Bianchi identities), it is possible  
to define more than one notion of  charge associated with a given gauge
field. Let us discuss here, following the presentation of ref.\cite{MarolfCB}, two particular definitions of this quantity, namely the so-called Maxwell and Page charges \cite{Page}.
Given a gauge invariant field strength $F_{8-p}$, the
(magnetic) Maxwell current associated to it is defined through  the following
relation:
\beq
d\,F_{8-p}\,=\,\star j^{Maxwell}_{D_{p}}\,\, ,
\label{Maxwell-j}
\eeq
or equivalently, the Maxwell charge in a volume $V_{9-p}$ is
 given by:
\beq
Q^{Maxwell}_{D_p}\,\sim \, \int_{V_{9-p}} \star j^{Maxwell}_{D_{p}}\,\, ,
\eeq
with a suitable normalization. Taking $\partial V_{9-p}=M_{8-p}$ and using 
(\ref{Maxwell-j}) and Stokes theorem,  we can rewrite the previous expression as:
\beq
Q^{Maxwell}_{D_p}\,\sim  \int_{M_{8-p}}\,\,F_{8-p}\,\,.
\eeq

This notion of current is gauge invariant and conserved and it has 
other properties that are discussed in \cite{MarolfCB}. In particular, it
is not ``localized" in the sense that for a solution of pure supergravity
(for which $d\,F_{8-p}=H_3 \wedge F_{6-p}$) this current does not vanish.
These are the kind of charges we have calculated so far \eqref{Neff}-\eqref{Meff}, namely:
\bear
&&Q^{Maxwell}_{D5}\,=\,M_{eff}\,=\, 
{1 \over {4 \pi^2}} \int F_3 \, \, , \rc
&&Q^{Maxwell}_{D3}\,=\, N_{eff}\,=\, 
{1 \over {(4 \pi^2)^2}} \int F_5 \,\, .
\eear
An important issue regarding these charges is that, in general, they
are not quantized. Indeed, we have  checked explicitly  that
$Q^{Maxwell}_{D5}=M_{eff}$ and 
$Q^{Maxwell}_{D3}=N_{eff}$ vary continuously with the holographic
variable $r$ (see eqs. (\ref{Meff}) and (\ref{Neff})).

Let us move on to the notion of Page charge. The idea is first to write 
the Bianchi identities for $F_3$  and $F_5$ 
as the exterior derivatives of some differential
form, which in general will not be gauge invariant, and then introduce the
Page current as a source.  In  our
case, we can define the following (magnetic) Page currents:
\begin{equation}
\begin{split}
&d(F_3\,-\,B_2 \wedge F_1)\,=\, \star j^{Page}_{D5}\,\, , \\
&d(F_5\,-\,B_2 \wedge F_3 \,+\,{1 \over 2} B_2 \wedge B_2 \wedge F_1)\,=\, 
\star j^{Page}_{D3}\,\, .
\end{split} \label{Jpage}
\end{equation}
Notice that the
currents defined by the previous expression are ``localized" as a
consequence of the Bianchi identities satisfied by $F_3$ and $F_5$, namely
$dF_3\,=\,H_3\wedge F_1$ and $dF_5\,=\,H_3\wedge F_3$.
The Page charges $Q^{Page}_{D5}$ and $Q^{Page}_{D3}$ are just  defined
as the integrals of $\star j^{Page}_{D5}$ and $\star j^{Page}_{D3}$
with the appropriate normalization,  \ie:
\begin{equation}
\begin{split}
Q^{Page}_{D5}\,&=\,  {1 \over {4 \pi^2}} \int_{V_4} \star j^{Page}_{D5} 
\,\, , \\
Q^{Page}_{D3}\,&=\, {1 \over {(4 \pi^2)^2}} \int_{V_6} \star j^{Page}_{D3}\,\,,
\end{split}
\end{equation}
where $V_4$ and $V_6$ are submanifolds in the transverse space to the D5- and D3-branes respectively, which enclose the branes. By using the expressions of the currents $\star j^{Page}_{D5} $ and $\star j^{Page}_{D3}$ given in (\ref{Jpage}),  and by applying Stokes
theorem, we get:
\begin{equation}\label{D5D3Page}
\begin{split}
Q^{Page}_{D5}\,&=\,  {1 \over {4 \pi^2}} \int_{S^3} 
\Big(F_3\,-\,B_2 \wedge F_1\Big)\,\, , \\
Q^{Page}_{D3}\,&=\, {1 \over {(4 \pi^2)^2}} \int_{\mathcal{M}_5} \Big(\,
 F_5\,-\,B_2 \wedge F_3 \,+\,{1 \over 2} B_2 \wedge B_2 \wedge F_1
 \,\Big)\,\,,
\end{split}
\end{equation}
where $S^3$ and $\mathcal{M}_5$ are the same manifolds used to compute the
Maxwell charges in eqs. (\ref{Meff}) and (\ref{Neff}).  It is not difficult to establish the topological nature of these Page charges. Indeed, let us consider, for concreteness, the expression of $Q^{Page}_{D5}$ in (\ref{D5D3Page}). Notice that the three-form under the integral can be {\it locally} represented as the exterior derivative of a two-form, since
$F_3-B_2\wedge F_1=dC_2$, with $C_2$ being the RR two-form potential.  If $C_2$ were well-defined globally on the $S^3$, the Page charge $Q^{Page}_{D5}$ would vanish identically as a consequence of Stokes theorem. Thus, $Q^{Page}_{D5}$  can be naturally interpreted as a monopole number and it can be non-vanishing only in the case in which the gauge field is topologically non-trivial. For the D3-brane Page charge 
$Q^{Page}_{D3}$ a similar conclusion can be reached. 

Due to the topological nature of the Page charges defined above, one naturally expects that they are quantized and, as we shall shortly verify, they are independent of the holographic coordinate. This shows that they are the natural objects to compare with the numbers of branes that create the geometry in these backgrounds with varying flux. 
However, as it is manifest from the fact that $Q^{Page}_{D5}$ and $Q^{Page}_{D3}$ are given in (\ref{D5D3Page}) in terms of the $B_2$ field and not in terms of its  field strength $H_3$, the Page charges are not gauge invariant. In subsection \ref{Large-gauge} we will relate this non-invariance to the Seiberg duality of the field theory dual.

Let us now calculate the associated Page charges  for our ansatz (\ref{ansatz}) . We shall start by computing the D5-brane Page charge for the three-sphere $S^3$
defined by $\theta_2, \varphi_2={\rm constant}$. We already know  the value of the
integral of $F_3$, which gives precisely $M_{eff}$ (see eq. (\ref{Meff})). Taking into account that
\beq
\int_{S^3}\,g^5\wedge g^1\wedge g^2\,=\,\int_{S^3}
g^5\wedge g^3\wedge g^4\,=\,8\pi^2\,\,, \label{intS3}
\eeq
we readily get:
\beq
{1\over 4\pi^2}\,\int_{S^3}\,B_2\wedge F_1\,=\,{MN_f\over 4\pi}\,\,
(f+k)\,\,,
\eeq
and therefore:
\beq
Q^{Page}_{D5}\,=\,M_{eff}\,-\,{MN_f\over 4\pi}\,\,
(f+k)\,\,.
\label{QD5-Meff}
\eeq
Using the expression of $M_{eff}$ given in (\ref{Meff}), we obtain:
\beq
Q^{Page}_{D5}\,=\, M\,\,,
\label{QD5=M}
\eeq
which is certainly quantized and independent of the radial coordinate.

Let us now look at the D3-brane Page charge, which can be computed as an integral  over the angular manifold $M_5$. Taking into account that
\beq
\int_{\mathcal{M}_5}\,g^1\wedge g^2\wedge g^3\wedge g^4\wedge g^5\,=\,(4\pi)^3\,\,,
\label{int5g}
\eeq
we get that, for our ansatz (\ref{ansatz}):
\begin{equation}
\begin{split}
&{1 \over {(4 \pi^2)^2}} \int_{\mathcal{M}_5} B_2\wedge F_3\,=\,
{M^2\over \pi}\,\,\Big[\,f\,-\,(f-k)\,F\,+\,{N_f\over 2\pi}\,fk\,\Big]
\,\,,\\
&{1 \over {(4 \pi^2)^2}} \int_{\mathcal{M}_5} B_2\wedge B_2\wedge F_1\,=\,
{M^2\over \pi}\,{N_f\over 4\pi}\,fk\,\,,
\end{split}
\end{equation}
and, thus
\beq
Q^{Page}_{D3}\,=\,N_{eff}\,-\,
{M^2\over \pi}\,\,\Big[\,f\,-\,(f-k)\,F\,+\,{N_f\over 4\pi}\,fk\,\Big]\,\,,
\label{QD3-Neff}
\eeq
and, using the expression of $N_{eff}$, we obtain
\beq
Q^{Page}_{D3}\,=\,N_0\,\,,
\label{QD3=N0}
\eeq
which is again independent of the holographic coordinate. 
Recall that these Page charges are not gauge invariant and we will study in 
subsection \ref{Large-gauge} how they change under a large gauge transformation.

We now proceed to present the solutions to the BPS equations of motion.

\section{Flavored warped deformed conifold} \label{KSflavored}

Let us now consider the following solution of the algebraic constraint (\ref{g-constraint}):
\begin{equation}
\label{g}
g^2\,=\,1\,-\,e^{2(G_1-G_2)} \,\,.
\end{equation}
In order to write the equations for the metric and dilaton in this case, let us 
perform the following change of variable:
\begin{equation}
3\, e^{-G_3}\,dr\,=\,d\tau\,\, . 
\end{equation}
In terms of this new variable, the differential equation for the 
dilaton is simply:
\begin{equation}
\dot{\phi}\,=\,{N_f\over 4\pi}\,e^{\phi}\,\,,
\label{dotphi}
\end{equation}
where the dot means derivative with respect to $\tau$. This equation can be
straightforwardly integrated, namely:
\begin{equation}
{N_f\over 4\pi}\,e^{\phi}\,=\,{1\over \tau_0-\tau}\,\,,
\qquad\qquad
0\le \tau\le \tau_0\,\,,
\label{phi(tau)}
\end{equation}
where $\tau_0$ is an integration constant. 
Let us now write the equations imposed by supersymmetry  to the metric functions $G_1$, $G_2$
and $G_3$, which are:
\bear
&&\dot{G}_1\,-\,{1 \over 18}e^{2G_3-G_1-G_2}\,-\,{1 \over 2}
e^{G_2-G_1}\,+\,{1 \over 2}e^{G_1-G_2}\,=\,0\,\, , \rc
&&\dot{G_2}\,-\,{1 \over 18}e^{2G_3-G_1-G_2}\,+\,{1 \over 2}
e^{G_2-G_1}\,-\,{1 \over 2}e^{G_1-G_2}\,=\,0\,\, , \rc
&&\dot{G_3}\,+\,{1 \over 9}e^{2G_3-G_1-G_2}\,-\, e^{G_2-G_1}\,
+\,{N_f \over 8\pi}e^{\phi}\,=\,0\,\,.
\eear
In order to write the solution of this system of equations, let us 
define the following function
\begin{equation}
\Lambda(\tau)\,\equiv\,{
\Big[\,2(\tau-\tau_0)(\tau-\sinh 2\tau)\,+\,\cosh
(2\tau)\,-\,2\tau\tau_0\, -\,1\,\Big]^{{1\over 3}}\over
\sinh\tau}\,\,.
\end{equation}
Then, the metric functions $G_i$ are given by:
\bear
&&e^{2G_1}\,=\,{1\over  4}\,\,\mu^{{4\over 3}}\,
{\sinh^2\tau\over \cosh\tau}\,\Lambda(\tau)\,\,,
\qquad\qquad
e^{2G_2}\,=\,{1\over 4}\,\,\mu^{{4\over 3}}\,
\cosh\tau\,\Lambda(\tau)\,\,,\rc\rc
&&e^{2G_3}\,=\,6\,\mu^{{4\over 3}}\,\,{\tau_0-\tau\over
\big[\,\Lambda(\tau)\,\big]^2}\,\,,
\label{G-sol}
\eear
where $\mu$ is an integration constant. Notice that the range of 
$\tau$ variable chosen in (\ref{phi(tau)}) is the one that makes the dilaton and the metric
functions real. Moreover, for the solution we have found,
the  fibering function $g$ is given by:
\begin{equation}
g\,=\,{1\over \cosh\tau}\,\,.
\end{equation}
By using this result, we can write the metric as:
\bear  \label{metric2}
&&ds^2\,=\,\Big[\,h(\tau)\,\Big]^{-{1\over2}}\,dx^2_{1,3}\,+
\,\Big[\,h(\tau)\,\Big]^{{1\over2}}\,ds^2_{6}\,\,,
\eear
where $ds^2_{6}$ is the metric of the `flavored' deformed conifold, namely
\bear
&&ds^2_{6}\,=\,{1\over 2}\,\,\mu^{{4\over 3}}\,\,\Lambda(\tau)\,\,
\Bigg[\,{4(\tau_0-\tau)\over 
3\Lambda^3(\tau)}\,\,\big(\,d\tau^2\,+\,(g^5)^2\,\big)\,+\,
  \cosh^2\Big({\tau\over 2}\big)\,\Big(\,(g^3)^2\,+\, 
(g^4)^2\,\Big)\,+\,\,\rc
  &&\qquad\qquad\qquad\qquad\qquad\qquad\qquad
  +\,\sinh^2\Big({\tau\over 2}\Big)\,
\Big(\,(g^1)^2\,+\, (g^2)^2\,\Big)\,
\,\Bigg]\,\,.
\label{flav-def-metric}
\eear
Notice the similarity between the metric (\ref{flav-def-metric}) and the one corresponding to the `unflavored' deformed conifold \cite{Klebanov:2000hb}. To further analyze this similarity, let us study the $N_f\to 0$ limit of our solution.  By looking at the expression of the dilaton in (\ref{phi(tau)}), one realizes that this limit is only sensible if one also sends $\tau_0\to+\infty$ with 
$N_f\tau_0$ fixed. Indeed, by performing this scaling and neglecting $\tau$ versus 
$\tau_0$, one gets a constant value for the dilaton. Moreover, the function $\Lambda(\tau)$ reduces in this limit to $\Lambda(\tau)\approx (4\tau_0)^{{1\over 3}}\,\,K(\tau)$, where $K(\tau)$ is the function appearing in the metric of the deformed conifold, namely:
\begin{equation}
K(\tau)\,=\,{\Big[\,\sinh 2\tau-
2\tau\,\Big]^{{1\over 3}}\over 2^{{1\over 3}}
\sinh\tau}\,\,.
\end{equation}
By using this result one easily verifies that, after redefining 
$\mu\to\mu/(4\tau_0)^{{1\over 4}}$, the metric (\ref{flav-def-metric}) reduces to the one used in  \cite{Klebanov:2000hb} for the unflavored system.

The requirement of supersymmetry imposes the following differential equations for the functions $k$,
$f$ and $F$ appearing in the fluxes of our ansatz:
\bear
&&\dot{k}\,=\,e^{\phi}\,\Big(\,F\,+\,{N_f\over 4\pi}\,f\,\Big)\,
\coth^2{{\tau\over 2}}\,\,,\rc
&&\dot{f}\,=\,e^{\phi}\,\,\Big(\,1\,-\,F\,+\,{N_f\over 4\pi}\,k\,\Big)\,
\tanh^2{\tau\over 2}\,\,,  \rc
&&\dot{F}\,=\,{1 \over 2}e^{-\phi} (k-f)\,\,.
\label{kfF-with-tau}
\eear
Notice, again, that for $N_f=0$ the system (\ref{kfF-with-tau}) reduces to the one 
found in \cite{Klebanov:2000hb}. Moreover, for $N_f\not=0$ this system can be solved as:
\bear
\label{sol}
&&e^{-\phi}\,f\,=\,{{\tau \coth{\tau}-1} \over {2 \,
\sinh{\tau}}}(\cosh{\tau}-1)\,\,,
\qquad\qquad
e^{-\phi}\,k\,=\,{{\tau \coth{\tau}-1} \over {2 \, \sinh{\tau}}}
(\cosh{\tau}+1)\,\, , \rc
&&\qquad\qquad\qquad\qquad\qquad\qquad
F\,=\,{{\sinh{\tau}-\tau} \over {2\, \sinh{\tau}}}\,\,,
\eear
where $e^{\phi}$ is given in eq. (\ref{phi(tau)}).
By using the solution given by (\ref{G-sol}) and (\ref{sol}) in the
general eq.  (\ref{Kh2-fkF}) we can immediately obtain the expression
of the warp factor $h(\tau)$.  Actually, if we require that $h$ is
regular at $\tau=0$, the integration constant $N_0$ in \eqref{Neff} must
be chosen to be zero. In this case, we get:
\begin{equation}
h(\tau)=-{{\pi \, M^2} \over {4\,\mu^{8/3} N_f}}
\int^{\tau} dx {{x \coth{x}-1} \over {(x\,-\,\tau_0)^2 \sinh^2{x}}}
{{-\cosh{2x}\,+\,4x^2\,-\,4x\tau_0\,+\,1\,-\,(x\,-\,2\tau_0)\sinh{2x}} \over
{(\cosh{2x}\,+\,2x^2\,-\,4x\tau_0\,-\,1\,-\,2(x\,-\,\tau_0)\sinh{2x})^{2/3}}} 
\,\, .
\end{equation}
The integration constant can be fixed by requiring that the analytic continuation of $h(\tau)$ goes to zero as $\tau \to\ +\infty$, to connect with the Klebanov-Strassler solution in the unflavored (scaling) limit. Then, close to the tip of the geometry, $h(\tau) \sim h_0 - \mathcal{O}(\tau^2)$.

We should emphasize now an important point: even though at first sight this solution may look smooth in the IR ($\tau\sim 0$), where all the components of our metric approach the same limit as those of the KS solution (up to a suitable redefinition of parameters),  there is actually a curvature singularity.
Indeed, in Einstein frame the curvature scalar behaves as $R_E \sim 1/\tau$.%
\footnote{The simplest example of this kind of singularity appears at $r=0$ in a 2-dimensional  manifold whose metric is $ds^2= dr^2 + r^2 (1+r)d\varphi^2$.}
This singularity of course disappears when taking the unflavored limit, using the scaling described above. 

The solution presented above is naturally interpreted as the addition of fundamentals to the KS background \cite{Klebanov:2000hb}. In the next section, we will present a solution that can be understood as the addition of flavors to the KT background \cite{Klebanov:2000nc}.

\section{Fractional branes in the singular conifold with flavor} \label{KTflavored}

Let us now consider the solutions with $g=0$. First of all, let us change
the radial variable from $r$ to $\rho$, where the later is defined by the
relation  $dr=e^{G_3}\,d\rho$. The equation for the dilaton can be
integrated trivially:
\begin{equation}
e^{\phi}\,=\,-{4\pi\over 3N_f}\,\,{1\over \rho}\,\,,
\qquad\qquad \rho<0\,\,.
\end{equation}
The supersymmetry requirement imposes now that
the metric functions $G_i$ satisfy  in this case the following system of
differential equations:
\bear
&&\dot G_i\,=\,{1\over 6}\,e^{2G_3-2G_i}\,\,,
\qquad\qquad (i=1,2)\,\,,\rc\rc
&&\dot G_3\,=\,3\,-\,{1\over 6}\,e^{2G_3-2G_1}
\,-\,{1\over 6}\,e^{2G_3-2G_2}\,-\,{3N_f\over 8\pi}\,e^{\phi}\,\,,
\eear
where now the dot refers to the derivative with respect to $\rho$.
This system is equivalent to the one analyzed in \cite{Benini:2006hh} for the
Klebanov-Witten model with flavors. In what follows we will restrict
ourselves to the particular  solution with $G_1=G_2$ given by:
\begin{equation}
e^{2G_1}\,=\,e^{2G_2}\,=\,{1\over 6}\,(1-6\rho)^{{1\over 3}}\,
e^{2\rho}\,\,,
\qquad\qquad
e^{2G_3}\,=\,-6\rho\,(1-6\rho)^{-{2\over 3}}\,e^{2\rho}\,\,.
\end{equation}
Notice that, as in \cite{Benini:2006hh}, the range of values of $\rho$ for
which the metric is well defined is $-\infty<\rho<0$. The equations for
the flux functions  $f$, $k$  and $F$ are now:
\bear
&&\dot f\,-\,\dot k\,=\,2e^{\phi}\dot F\,\,,\rc\rc
&&\dot f\,+\,\dot k\,=\,3e^{\phi}
\Big[\,1\,+\,{N_f\over 4\pi}(f+k)\,\Big]\,\,,\rc\rc
&&F\,=\,{1\over 2}\,\Big[\,1\,+\,
\Big(\,e^{-\phi}\,-\,{N_f\over 4\pi}\,\Big)\,
(f-k)\,\Big]\,\,.
\eear
In this paper, we will focus on the particular solution of this system
such that $f=k$ and $F$ is constant, namely:
\begin{equation}
F={1\over 2}\,\,,
\qquad\qquad
f\,=\,k\,=\,-{2\pi\over N_f}\,\Bigg(\,1\,-\,
{\Gamma\over \rho}\,\Bigg)\,\,,
\label{KT-Ffk-sol}
\end{equation}
where $\Gamma$ is an integration constant. By substituting these values
of $F$, $f$ and $k$ in our ansatz (\ref{ansatz}) we obtain the form of
$F_3$ and $H$. Notice that the constants $M$ and $\Gamma$ only appear 
in the combination
$M\Gamma$. Accordingly, let us define ${\cal M}$ as
${\cal M}\,\equiv\,M\Gamma$. We will write the result in terms of the
function:
\begin{equation}
M_{eff}(\rho)\equiv {{\cal M}\over \rho}\,\,.
\end{equation}
One has:
\begin{equation}
\begin{split}
F_3 \,&=\,{M_{eff}(\rho)\over 4}\,\,g^5\wedge
\big(g^1\wedge g^2\,+\,g^3\wedge g^4\big)\,\,,\\
H_3\,&=\,-{\pi\over N_f}\,{M_{eff}(\rho)\over \rho}\,\,d\rho\wedge
\big(g^1\wedge g^2\,+\,g^3\wedge g^4\big)\,\,.
\end{split}\label{F3 H3 KT}
\end{equation}
Moreover, the RR five-form $F_5$ can be written as in (\ref{F5-Neff}) in
terms of the effective D3-brane charge defined in \eqref{Neff}. For the
solution  (\ref{KT-Ffk-sol}) one gets:
\begin{equation}
N_{eff}(\rho)\,=\,N\,+\,{{\cal M}^2\over N_f}\,{1\over \rho^2}\,\,,
\end{equation}
where  $N\,\equiv\,N_0\,-\,{ M^2\over N_f}$. By using eq.
(\ref{Kh2-fkF}), one can obtain the expression of the warp factor, namely:
\begin{equation}
h(\rho)\,=\,-27\pi \int\,d\rho\,
\Bigg[\,N\,+\,{{\cal M}^2\over N_f}\,{1\over \rho^2}\,\Bigg]\,
{e^{-4\rho}\over (1-6\rho)^{{2\over 3}}}\,\,.
\end{equation}
To interpret the solution just presented, it is interesting to study it in the deep IR region 
$\rho\to-\infty$. Notice that in this limit the three-forms $F_3$ and $H_3$ vanish. Actually, it is easy to verify that for $\rho\to-\infty$ the solution obtained here reduces to the one studied in \cite{Benini:2006hh}, corresponding to the Klebanov-Witten 
\cite{Klebanov:1998hh} model with flavors. Indeed, in this IR region it is convenient to go back to our original radial variable $r$. The relation between $r$ and $\rho$ for $\rho\to-\infty$ is $r\approx (-6\rho)^{{1\over 6}}\,e^{\rho}$. Moreover, one can prove that for 
$\rho\to-\infty$  (or equivalently $r\to 0$), the warp factor $h$ and the metric functions $G_i$ become:
\begin{equation}
h(r)\approx {27\pi N\over 4}\,{1\over r^4}\,\,,
\qquad\qquad
e^{2G_1}\,=\,e^{2G_2}\approx {r^2\over 6}\,\,,
\qquad\qquad
e^{2G_3}\,\approx\,r^2\,\,,
\end{equation}
which implies that the IR Einstein frame metric is $AdS_5\times T^{1,1}$ plus logarithmic corrections, exactly as the solution found in \cite{Benini:2006hh}.
The interpretation of the RG flow of the field theory dual to this solution will be explained in Sections \ref{sect: Field Th} and \ref{cascade-SUGRA}.

Finally, let us stress that the UV behavior of this solution (coincident with that of the solution presented in Section \ref{KSflavored}) presents a divergent dilaton at the point $\rho=0$ (or $\tau=\tau_0$ for the flavored warped deformed conifold). Hence the supergravity approximation fails at some value of the radial coordinate that we will associate in Section \ref{cascade-SUGRA} with the presence of a duality wall \cite{Strassler:1996ua} in the cascading field theory.

\section{The field theory with flavors: a cascade of Seiberg dualities} \label{sect: Field Th}

The field theory dual to our supergravity solutions can be engineered by putting stacks of two kinds of fractional D3-branes (color branes) and two kinds of fractional D7-branes (flavor branes) on the singular conifold.
The smeared charge distribution introduced in the previous sections can be realized by homogeneously distributing D7-branes among a class of localized $\kappa$-symmetric embeddings. The (complex structure of the) deformed conifold is described by one equation in $\bbC^4$: $z_1 z_2 - z_3 z_4 = \epsilon^2$. This has isometry group $SU(2)_A\times SU(2)_B$, where the non-abelian factors are realized through left and right multiplication on the matrix $\bigl( \begin{smallmatrix} z_1 & z_4 \\ z_3 & z_2 \end{smallmatrix} \bigr)$. We can also define a $U(1)_R$ action, which is a common phase rotation, that is broken to $\bbZ_2$ by the deformation parameter $\epsilon$. Consider the embedding \cite{Kuperstein:2004hy}:
\begin{equation} \label{embedding eq}
z_3 + z_4 = 0 \;.
\end{equation}
This is invariant under $U(1)_R$ and a diagonal $SU(2)_D$ (and a $\bbZ_2$ which exchanges $z_3 \leftrightarrow z_4$). Moreover it is free of $C_8$ tadpoles and it was shown to be $\kappa$-symmetric in \cite{Kuperstein:2004hy}. It could be useful to write it in the angular coordinates of the previous section: $\theta_1=\theta_2$, $\varphi_1 = \varphi_2$, $\forall \psi,\forall\tau$. We can obtain other embeddings with the same properties by acting on it with the broken generators. One can show that the charge distribution obtained by homogeneously spreading the D7-branes in this class is \eqref{smearing}:
\begin{equation}
\Omega = \frac{N_f}{4\pi} \bigl( \sin\theta_1 \, d\theta_1\wedge d\varphi_1 + \sin\theta_2 \, d\theta_2 \wedge d\varphi_2 \bigr) \;,
\end{equation}
where $N_f$ is the total number of D7-branes.

Notice that one could have considered the more general embedding: $z_3 + z_4 = m$, where $m$ corresponds in field theory to a mass term for quarks. These embeddings and their corresponding supergravity solutions are not worked out in this paper.

\begin{figure}[ht]
\begin{center}
\includegraphics[width=0.8\textwidth]{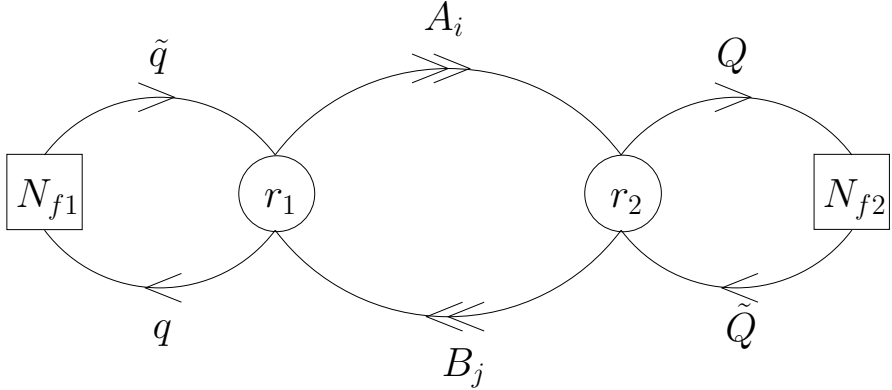}
\end{center}
\caption[quiverN]{The quiver diagram of the gauge theory. Circles are gauge groups, squares are flavor groups, and arrows are bifundamental chiral superfields. $N_{f1}$ and $N_{f2}$ sum up to $N_f$. \label{quiverN}} 
\end{figure}

Different techniques have been developed to identify the field theory dual to our Type IIB plus D7-branes background, which can be engineered by putting $r_l$ fractional D3-branes of the first kind, $r_s$ fractional D3-branes of the second kind, $N_{fl}$ fractional D7-branes of the first kind, and $N_{fs}$ fractional D7-branes of the second kind ($l,s=1,2$) on the singular conifold, before the deformation has dynamically taken place. The properties of the different kinds of fractional branes will be explained at the end of this Section and in Section \ref{cascade-SUGRA}; what matters for the time being is that this brane configuration will give rise to a field theory with gauge groups $SU(r_l)\times SU(r_s)$ and flavor groups $SU(N_{fl})$ and $SU(N_{fs})$ for the two gauge groups respectively, with the  matter content displayed in Fig. \ref{quiverN}.
The most convenient technique for our purpose has been that of performing a T-duality along the isometry $(z_1,z_2)\to(e^{i\alpha}z_1,e^{-i\alpha}z_2)$ (one does not need the metric, only the complex structure). The system is mapped into Type IIA: neglecting the common spacetime directions, there is a NS5-brane along $x^{4,5}$, another orthogonal NS5 along $x^{8,9}$, $r_l$ D4-branes along $x^6$ (which is a compact direction) connecting them on one side, other $r_s$ D4 connecting them on the other side, $N_{fl}$ D6-branes along $x^7$ and at a $\frac{\pi}{4}$ angle between $x^{4,5}$ and $x^{8,9}$, touching the stack of $r_l$ D4-branes, and $N_{fs}$ D6-branes along $x^7$ and at a $\frac{\pi}{4}$ angle between $x^{8,9}$ and $x^{4,5}$, touching the stack of $r_s$ D4-branes
. Then the spectrum is directly read off, and the superpotential comes from the analysis of the moduli space \cite{Radu}.

The field content of the gauge theory can be read from the quiver diagram of Figure \ref{quiverN}: it is an extension of the Klebanov-Strassler field theory with nonchiral flavors for each gauge group.
The superpotential is%
\footnote{Sums over gauge and flavor indices are understood.}
\begin{equation}
\begin{split}
W & = \lambda (A_1 B_1 A_2 B_2 -A_1 B_2 A_2 B_1) + h_1 \, \tilde{q} (A_1 B_1 + A_2 B_2)q + h_2 \, \tilde{Q} (B_1 A_1 + B_2 A_2)Q + \\
&\quad +\alpha \, \tilde{q} q \tilde{q} q + \beta \, \tilde{Q} Q \tilde{Q} Q  \;.
\end{split}
\end{equation}
The factors $A_1B_1 + A_2B_2$ directly descend from the embedding equation \eqref{embedding eq}, while the quartic term in the fundamental fields is derived from Type IIA.
This superpotential explicitly breaks the $SU(2)_A \times SU(2)_B$ global symmetry of the unflavored theory to $SU(2)_D$, but this global symmetry is recovered after the smearing (see \cite{Benini:2006hh} for a more careful treatment of the smearing procedure). It's worth here to stress that the smearing procedure does not influence at all either the duality cascade, which is the main feature of our solutions that we want to address here, nor (presumably) the infrared dynamics which will be investigated in a forthcoming paper \cite{us}.

The $N_f$ flavors are split into $N_{fl}$ and $N_{fs}$ groups, according to which gauge group they are charged under. Both sets come from D7-branes along the embedding \eqref{embedding eq}.%
\footnote{The embedding is in fact invariant under the $\bbZ_2$ ($z_3 \leftrightarrow z_4$) that exchanges the two gauge groups.} 
The only feature that discriminates between these two kinds of (fractional) D7-branes is their coupling to the $C_2$ and $C_4$ gauge potentials. On the singular conifold, before the dynamical deformation, there is a vanishing 2-cycle, living at the singularity, which the D7-branes are wrapping.%
\footnote{Since the D7-brane has 4 internal directions, even if it wraps the two-cycle living only at the singularity, still out of it is four-dimensional.} According to the worldvolume flux on it, the D7-branes couple either to one or the other gauge group. Since this flux is stuck at the origin, far from the branes we can only measure the D3, D5 and D7-charges produced. Unfortunately three charges are not enough to fix four ranks. This curious ambiguity will show up again in Section \ref{cascade-SUGRA}. 

\subsection{The cascade} \label{cascade subsection}

One can assume that, as in the unflavored case, the beta functions of the two gauge couplings have opposite sign. When the gauge coupling of the gauge group with larger rank is very large, one can go to a Seiberg-dual description \cite{Seiberg:1994pq}: remarkably, it's straightforward to see that the quartic superpotential is such that the field theory is self-similar, namely the field theory in the dual description is a quiver gauge theory with the same field content and superpotential, except for changes in the ranks of the groups.%
\footnote{This is not the case for the chirally flavored version of Klebanov-Strassler's theory proposed by Ouyang \cite{Ouyang:2003df}, and for the flavored version of nonconformal theories obtained by putting branes at conical Calabi-Yau singularities \cite{Franco:2006es}. In those realizations the superpotential is cubic, and the theory is not self-similar under  Seiberg duality: new gauge singlet fields appear or disappear after a Seiberg duality, making the cascade subtler.}

Let us define the theory at some energy scale to be an $SU(r_l)\times SU(r_s)$ gauge theory (where $l$ stands for the larger gauge group and $s$ for the smaller: $r_l > r_s$), with flavor group $SU(N_{fl})$ ($SU(N_{fs})$) for $SU(r_l)$ ($SU(r_s)$). In the beginning we can set, conventionally, $r_l=r_1$, $r_s=r_2$, $N_{fl}=N_{f1}$, $N_{fs}=N_{f2}$; after a Seiberg duality on the gauge group with the larger rank, the field theory is $SU(2r_2-r_1+N_{f1})\times SU(r_2)$, with again $N_{f1}$ and $N_{f2}$ flavors respectively. In identifying which gauge group is now the larger and which is the smaller, we have to exchange the labelling of the groups, so that we get  $r'_l=r_2$, $r'_s=2r_2-r_1+N_{f1}$, $N'_{fl}=N_{f2}$ and $N'_{fs}=N_{f1}$.
The assumption leads to an RG flow which is described by a cascade of Seiberg dualities, analogous to \cite{Klebanov:2000nc,Klebanov:2000hb}. In the UV the ranks of the gauge groups are much larger than their disbalance, which is much larger than the number of flavors. Hence the assumption of having beta functions with opposite sign is justified in the ultraviolet flow of the field theory.

The supergravity background of Section \ref{KSflavored} is dual to a quiver gauge theory where the cascade goes on until the IR, with nonperturbative dynamics at the end, as in the Klebanov-Strassler solution. 

In the background of Section \ref{KTflavored}, the cascade does not take place anymore below some value of the radial coordinate, and it asymptotes to the flavored Klebanov-Witten solution \cite{Benini:2006hh}. In the field theory, this reflects the fact that, because of a suitable choice of the ranks, the last step of the cascade leads to a theory where the beta functions of both  gauge couplings are positive. The infrared dynamics is the one discussed in \cite{Benini:2006hh}, but with a quartic superpotential for the flavors.

The description of the duality cascade in our solutions and its interesting ultraviolet behavior will be the content of the next section.

\section{The cascade: supergravity side}\label{cascade-SUGRA}

We claim that our supergravity solutions are dual to the class of quiver gauge theories with backreacting fundamental flavors introduced in the previous section. Indeed we will show that the effective brane charges, the R-anomalies and the beta functions of the gauge couplings that we can read from the supergravity solutions precisely match the picture of a cascade of Seiberg dualities that we expect to describe the RG flow of the field theories, generalizing the results of \cite{Klebanov:2000nc,Klebanov:2000hb} to gauge theories which include dynamical flavors.

\subsection{Effective brane charges and ranks}
\label{charges and ranks}

By integrating fluxes over suitable compact cycles, we can compute three effective D-brane charges in our solutions, which are useful to pinpoint the changes in the ranks of gauge  groups when the field theory undergoes a Seiberg duality: one of them (D7) is dual to a quantity which is constant along the RG flow, whereas two of them (D3, D5) are not independent of the holographic coordinate and are  dual to the nontrivial part of the RG flow. Recall that the (Maxwell) charges of D3- and D5-brane ($N_{eff}$ and $M_{eff}$)  for our ansatz were  already calculated in section \ref{section-ansatz} (see eqs. (\ref{Neff}) and (\ref{Meff})). 
Let us now compute the D7-brane charge, integrating \eqref{smearing} on a 2-manifold with boundary   which is intersected once by all the smeared D7-branes (e.g. $\mathcal{D}_2$: $\theta_2=\text{const.}$, $\varphi_2=\text{const.}$, $\psi=\text{const.}$). This charge is conserved along the RG flow because no fluxes appear on the right hand side of \eqref{smearing}. The D7-brane charge, which we interpret as the total number of flavors added to the Klebanov-Strassler gauge theory, is indeed:
\begin{equation} \label{Nflav}
N_{flav}\equiv \int_{\mathcal{D}_2} dF_1 = N_f\;.
\end{equation}
Another important quantity is the integral of $B_2$ over the nontrivial 2-cycle $S^2$: $\theta_1 = \theta_2 \equiv \theta$, $\varphi_1 = 2\pi -\varphi_2 \equiv \varphi$, $\psi=\text{const.}$:
\begin{equation} \label{b_0}
b_0(\tau)\equiv \frac{1}{4\pi^2} \int_{S^2} B_2 =  \frac{M}{\pi} \Bigl( f \sin^2\frac{\psi}{2} + k\cos^2\frac{\psi}{2} \Bigr)\;.
\end{equation}
This quantity is important because string theory is invariant as it undergoes a shift of 1. For instance, in the KW background it amounts to a Seiberg duality, and the same happens here.
So we will shift this last quantity by one unit, identify a shift in the radial variable $\tau$ that realizes the same effect, and see what happens to $M_{eff}$ and $N_{eff}$. Actually, the cascade will not work along the whole flow down to the IR but only in the UV asymptotic (below the UV cut-off $\tau_0$ obviously). Notice that the same happens for the unflavored solutions of \cite{Klebanov:2000nc} and \cite{Klebanov:2000hb}: in the KT solution one perfectly matches the cascade in field theory and supergravity, while in the KS solution close to the tip of the warped deformed conifold the matching is not so clean. On the other hand, this is expected, since the last step of the cascade is not a Seiberg duality. Thus we will not be worried and compute the cascade only in the UV asymptotic for large $\tau$ which also requires $\tau_0 \gg 1$ (we neglect $\mathcal{O}(e^{-\tau})$): in that regime the functions $f$ and $k$ become equal, and $b_0$ is $\psi$-independent.\\
Actually, we will not compute the explicit shift in $\tau$ but rather the shift in the functions $f$ and $k$.
We have:
\begin{equation}
\begin{split}
b_0 (\tau) \to b_0(\tau') = b_0 (\tau) - 1 
\end{split}
\quad 
\Longrightarrow \quad 
\begin{split}
f(\tau) &\to f(\tau') = f(\tau)  - \frac{\pi}{M} \\
k(\tau) &\to k(\tau') = k(\tau)  - \frac{\pi}{M} 
\end{split} \label{shift f k}
\end{equation} 
Correspondingly, after a Seiberg duality step from $\tau$ to $\tau'<\tau$, that is going towards the IR, we have:
\begin{align} 
N_f &\to N_f \label{scaling sugra 1} \\
M_{eff} (\tau) &\to M_{eff} (\tau') = M_{eff} (\tau) - \frac{N_f}{2} \label{scaling sugra Meff}\\
N_{eff} (\tau) &\to N_{eff} (\tau') = N_{eff} (\tau) - M_{eff} (\tau) + \frac{N_f}{4} \label{scaling sugra 2}
\end{align}
This result is valid for all of our solutions.

We would like to compare this result with the action of Seiberg duality in field theory, as computed in Section \ref{sect: Field Th}. We need an identification between the brane charges computed in supergravity and the ranks of the gauge and flavor groups in the field theory.

The field theory of interest for us, with gauge groups $SU(r_l) \times SU(r_s)$ ($r_l>r_s$), and flavor groups $SU(N_{fl})$ and $SU(N_{fs})$ for the gauge groups $SU(r_l)$ and $SU(r_s)$ respectively, is engineered, at least effectively at some radial distance, by the following objects: $r_l$ fractional D3-branes of one kind (D5-branes wrapped on the shrinking 2-cycle), $r_s$ fractional D3-branes of the other kind ($\overline{\text{D}5}$-branes wrapped on the shrinking cycle, supplied with $-1$ quanta of gauge field flux on the 2-cycle), $N_{fs}$ fractional D7-branes without gauge field strength on the 2-cycle, and $N_{fl}$ fractional D7-branes with $-1$ units of gauge field flux on the shrinking 2-cycle. This description is good for $b_0\in [0,1]$.

This construction can be checked explicitly in the case of the $\mathbb{C} \times \mathbb{C}^2/\mathbb{Z}_2  $ orbifold \cite{Bertolini:2000dk, Bertolini:2001qa}, where one is able to quantize the open and closed string system for the case $b_0=\frac{1}{2}$ that leads to a free CFT \cite{Aspinwall:1995zi}. That is the $\mathcal{N}=2$ field theory which flows to the field theory we are considering, when equal and opposite masses are given to the adjoint chiral superfields (the geometric description of this relevant deformation is a blowup of the orbifold singularity)  \cite{Klebanov:1998hh,Morrison:1998cs}. Fractional branes are those branes which couple to the twisted closed string sector.%
\footnote{Notice that one can build, out of a fractional D3 of one kind and a fractional D3 of the other kind, a regular D3-brane (\emph{i.e.} not coupled to the twisted sector) that can move outside the orbifold singularity; on the contrary, there is no regular D7-brane: the two kinds of fractional D7-branes, extending entirely along the orbifold, cannot bind into a regular D7-brane that does not touch the orbifold fixed locus and is not coupled to the twisted sector \cite{Bertolini:2001qa}.}

Here we will consider a general background value for $B_2$. In order to compute the charges, we will follow quite closely the computations in \cite{Grana:2001xn}.

We will compute the charges of D7-branes and wrapped D5-branes on the singular conifold, described by $z_1 z_2 - z_3 z_4 = 0$. The D5 Wess-Zumino action is
\begin{equation}
S_{D5} = \mu_5 \int_{M^4\times S^2} \Bigl\{ C_6 + (2\pi\, F_2 + B_2)\wedge C_4 \Bigr\}\;,
\end{equation} 
where $S^2$ is the only 2-cycle in the conifold, vanishing at the tip, that the D5-brane is wrapping. We write also a world-volume gauge field $F_2$ on $S^2$. Then we expand:
\begin{equation}
B_2 = 2\pi\, \theta_B\, \omega_2 \qquad\qquad \theta_B = 2\pi \, b_0 \qquad\qquad F_2 = \Phi\, \omega_2\;,
\end{equation}
where $\omega_2$ is the 2-form on the 2-cycle, which satisfies $\int_{S^2} \omega_2 = 1$. In this conventions, $b_0$ has period 1, and $\Phi$ is quantized in $2\pi \, \mathbb{Z}$. We obtain (using $\mu_p (4\pi^2)=\mu_{p-2}$):
\begin{equation}
S_{D5} = \mu_5 \int_{M^4\times S^2} C_6 + \frac{\mu_3}{2\pi} \int_{M^4} (\Phi + \theta_B) \, C_4 \;.
\end{equation}
The first fractional D3-brane \cite{Polchinski:2000mx} is obtained with $\Phi=0$ and has D3-charge $b_0$, D5-charge 1. The second fractional D3-brane is obtained either as the difference with a D3-brane, or as an anti-D5-brane (global $-$ sign in front) with $\Phi=-2\pi$, and has D3-charge $1-b_0$, D5-charge -1. These charges are summarized in Table \ref{Table:general charges}.

Now consider a D7-brane along the surface $z_3 + z_4 = 0$. It describes a $z_1 z_2 + z_3^2=0$ inside the conifold, which is a copy of $\mathbb{C}^2/\mathbb{Z}_2$.  The D7 Wess-Zumino action is (up to a curvature term considered below)
\begin{equation}
S_{D7} = \mu_7 \int_{M^4\times \Sigma} \Bigl\{ C_8 + (2\pi\, F_2 + B_2) \wedge C_6 + \frac{1}{2} (2\pi\, F_2 + B_2) \wedge (2\pi\, F_2 + B_2) \wedge C_4 \Bigr\}\;.
\end{equation}
The surface $\Sigma = \mathbb{C}^2/\mathbb{Z}_2$ has a vanishing 2-cycle at the origin. Since the conifold has only one 2-cycle, these two must be one and the same and we can expand on $\Sigma$ using $\omega_2$ again. Moreover, being $\omega_2$ the Poincar\'e dual to the 2-cycle on $\Sigma$, 
\begin{equation}
\int_\Sigma \omega_2 \wedge \alpha_2 = \frac{1}{2} \int_{S^2} \alpha_2
\end{equation} 
holds for any closed 2-form $\alpha_2$, and $\frac{1}{2}$ arises from the self-intersection number of the $S^2$. There is another contribution of induced D3-charge coming from the curvature coupling \cite{Bershadsky:1995qy}:
\begin{equation}
\frac{\mu_7}{96} (2\pi)^2 \int_{M^4\times \Sigma} C_4 \wedge \Tr \, \mathcal{R}_2\wedge \mathcal{R}_2 = -\mu_3 \int_{M^4\times \Sigma} C_4 \wedge \frac{p_1(\mathcal{R})}{48}\;.
\end{equation}
This can be computed in the following way. On K3 $p_1(\mathcal{R})= 48$ and the induced D3-charge is $-1$. In the orbifold limit K3 becomes $T^4/\mathbb{Z}_2$ which has 16 orbifold singularities, thus on $\mathbb{C}^2/\mathbb{Z}_2$ the induced D3-charge is $-1/16$. 
Putting all together we get:
\begin{equation}
S_{D7} = \mu_7 \int_{M^4\times \Sigma} C_8 + \frac{\mu_5}{4\pi} \int_{M^4 \times S^2} (\Phi + \theta_B) \, C_6 + \frac{\mu_3}{16\pi^2} \int_{M^4} \Bigl[ (\Phi + \theta_B)^2 - \pi^2 \Bigr] \, C_4\;.
\end{equation} 
The second fractional D7-brane (the one that couples to the second gauge group) is obtained with $\Phi=0$ and has D7-charge 1, D5-charge $\frac{b_0}{2}$ and D3-charge $(4b_0^2-1)/16$. With $\Phi=2\pi$ we get a non-SUSY or non-minimal object (see \cite{Polchinski:2000mx} for some discussion of this). The first fractional D7-brane (coupled to the first gauge group) has $\Phi=-2\pi$ and has D7-charge 1, D5-charge $\frac{b_0-1}{2}$ and D3-charge $(4(b_0-1)^2-1)/16$. This is summarized in Table \ref{Table:general charges}. 
Which fractional D7-brane provides flavors for the gauge group of which fractional D3-brane can be determined from the orbifold case with $b_0=\frac{1}{2}$ (compare with \cite{Bertolini:2001qa}).

\begin{table}[ht]
\begin{center}
\begin{tabular}{c|cccc}
 Object & frac D3 (1) & frac D3 (2) & frac D7 (1) & frac D7 (2) \\
\hline

D3-charge                    & $b_0$       & $1-b_0$     & $\dfrac{4(b_0-1)^2-1}{16}$ & $\dfrac{4b_0^2-1}{16}$ \\
\\
D5-charge                    & 1           & $-1$        & $\dfrac{b_0-1}{2}$ & $\dfrac{b_0}{2}$ \\
\\
D7-charge                    & 0           & 0           & 1           & 1\\
 \hline
 Number of objects & $r_l$ & $r_s$ & $N_{fl}$ & $N_{fs}$\\
\end{tabular}
\caption{Charges of fractional branes on the conifold \label{Table:general charges}}
\end{center}
\end{table}

Given these charges, we can compare with the field theory cascade. First of all we construct the dictionary:
\begin{align}
N_f &= N_{fl} + N_{fs} \\
M_{eff} &= r_l - r_s + \frac{b_0-1}{2} N_{fl} + \frac{b_0}{2} N_{fs}\label{dictionaryM} \\
N_{eff} &= b_0 \, r_l + (1-b_0) \, r_s + \frac{4(1-b_0)^2-1}{16} N_{fl} + \frac{4b_0^2-1}{16} N_{fs} \label{dictionary}
\end{align}
To derive this, we have only used that the brane configuration that engineers the field theory we consider consists of $r_l$ fractional D3 of the $1^{st}$ kind, $r_s$ fractional D3 of the $2^{nd}$ kind, $N_{fl}$ fractional D7 of the $1^{st}$ kind, and $N_{fs}$ fractional D7 of the $2^{nd}$ kind. Recall that, by convention, $r_l>r_s$ and $N_{fl}$ ($N_{fs}$) are the flavors for $SU(r_l)$ ($SU(r_s)$).\\
It is important to remember that $b_0$ is defined modulo 1, and shifting $b_0$ by one unit amounts to go to a Seiberg dual description in the field theory. At any given energy scale in the cascading gauge theory, there are infinitely many Seiberg dual descriptions of the field theory, because Seiberg duality is exact along the RG flow \cite{Strassler:2005qs}. Among these different pictures, there is one which gives the best effective description of the field theory degrees of freedom around that energy scale (this is also the  description with positive squared gauge couplings): it is the one where $b_0$ has been redefined, by means of a large gauge transformation, so that $b_0\in[0,1]$ (see subsection \ref{Large-gauge}). This is the description that we will use when we effectively engineer the field theory in terms of branes in some range of the RG flow that lies between two adjacent Seiberg dualities. \\
In field theory, as before, we start with gauge group $SU(r_1) \times SU(r_2)$ and $N_{f1}$ flavors for $SU(r_1)$, $N_{f2}$ flavors for $SU(r_2)$, with $r_1>r_2$. The gauge group $SU(r_1)$ flows towards strong coupling, and when its gauge coupling diverges we turn to a Seiberg dual description.
After the Seiberg duality on the larger gauge group,  we get $SU(2r_2 - r_1 + N_{f1}) \times SU(r_2)$, and the flavor groups are left untouched. \\
The effective D5- and D3-brane charges of a brane configuration that engineers this field theory {\it before} the duality are: 
\begin{equation}
\begin{split}
M_{eff} &= r_1-r_2+ \frac{b_0-1}{2}N_{f1} + \frac{b_0}{2}N_{f2}\;,\\
N_{eff} &= b_0 r_1 + (1-b_0)r_2 + \frac{4(1-b_0)^2-1}{16}N_{f1} + \frac{4b_0^2-1}{16}N_{f2}\;.
\end{split}
\end{equation} 
{\it After} the duality they become: 
\begin{equation}
\begin{split}
M'_{eff} &= -r_2 + r_1 - N_{f1} + \frac{b_0-1}{2}N_{f2} + \frac{b_0}{2}N_{f1}= M_{eff}-\frac{N_f}{2}\;,\\
N'_{eff} &= b_0r_2 + (1-b_0)(2r_2-r_1+N_{f1}) + \frac{4(1-b_0)^2-1}{16}N_{f2} + \frac{4b_0^2-1}{16}N_{f1}=\\ &=N_{eff}-M_{eff}+\frac{N_f}{4}\;.
\end{split}
\end{equation}
They \emph{exactly} reproduce the SUGRA behavior \eqref{scaling sugra 1}-\eqref{scaling sugra 2}.
Notice that the matching of the cascade between supergravity and field theory is there, irrespective of how we distribute the flavors between the two gauge groups; so, from the three charges and the cascade we are not able to determine how the flavors are distributed, but only their total number.
%

We conclude with some remarks. Even though the effective brane charges computed in supergravity are running and take integer values only at some values of the holographic coordinate, the ranks of gauge and flavor groups computed from them are constant and integer (for suitable choice of the integration constants) in the whole range of radial coordinate dual to the energy range where we use a specific field theory description. This range of scales is $b_0 \in [0,1]$ mod 1. At the boundaries of this region, we perform a Seiberg duality and go into a new more effective description, and in particular if ranks are integer before the duality, they still are after it; meanwhile we shift $b_0$ by one unit.
Hence the field theory description of the cascade is perfectly matched by the ranks that we can compute from our supergravity solution.

Notice also that the fact that $M_{eff}$ shifts by $N_f/2$ instead of $N_f$ confirms that the flavored version of the Klebanov-Strassler theory we are describing has nonchiral flavors (with a quartic superpotential) rather than chiral flavors (with a cubic superpotential) like in \cite{Ouyang:2003df}, where the shift goes with units of $N_f$. 

Finally, we want to stress again that we are engineering a field theory with 4 objects, but we have only 3 charges to recognize them. The comparison of the cascade between sugra and field theory, surprisingly enough, does not help.


\subsection{Seiberg duality as a large gauge transformation}
\label{Large-gauge}

We have argued in the previous subsection that a shift by a unit of the normalized flux $b_0$ 
as we move towards the IR along the holographic direction 
is equivalent to performing a Seiberg duality step on the field theory side (see equations \eqref{shift f k}-\eqref{scaling sugra 2}). Moreover, we have checked that, under this shift of $b_0$,  the change of the effective (Maxwell) charges $M_{eff}$ and $N_{eff}$ of supergravity is exactly the same as the one computed in the field theory engineered with fractional branes on the singular conifold. 

In this subsection we will present an alternative way of understanding in supergravity Seiberg duality at a {\it fixed} energy scale. As we know, for a given value of the holographic coordinate $\tau$, the value of $b_0$ lies generically outside the interval $[0,1]$, where a good field theory description exists. However, the flux of the $B_2$ field is not a gauge invariant quantity in supergravity and can be changed under a large gauge transformation. Indeed, let us define $\Upsilon_2$ as the following two-form:
\beq
\Upsilon_2\,=\, 
{1 \over 2} (g^1 \wedge g^2 \,+\, g^3 \wedge g^4) \,\, ,
\eeq
and let us change $B_2$ as follows: 
\beq
B_2\to B_2+\Delta B_2\,\,,
\qquad\qquad
\Delta B_2\,=\,-\pi n \Upsilon_2\,\,,
\qquad\qquad
n\,\in \mathbb{Z}\,\,.
\label{large}
\eeq
As $d\Upsilon_2=0$, the field strength $H_3$ does not change and our
transformation is a gauge  transformation of the NSNS field. However the
flux of $B_2$ does change as:
\beq
\int_{S^2}B_2 \to \int_{S^2}B_2 \,-\, 4 \pi^2 n \,\, ,
\label{flux-change}
\eeq
or, equivalently $b_0 \to b_0 - n$. This non-invariance of the flux shows
that this transformation of $B_2$ is a large gauge tranformation which
cannot be globally written as $\Delta  B_2=d\Lambda$. Moreover, as always
happens with large gauge transformations, it is
quantized. 
If we want that our transformation
(\ref{large})  be a gauge transformation of supergravity, it should leave the RR
field strength $F_3$ invariant. Defining the potential $C_2$ as
$dC_2=F_3-B_2 \wedge F_1 $, we see that $dC_2$ must change as:
\beq
dC_2 \to dC_2 \,+\, {{n N_f} \over 4} g^5 \wedge \Upsilon_2 \,\, .
\eeq
One can verify that this change of $dC_2$ can be obtained if the
variation of $C_2$ is (see equations \eqref{ansatz} and \eqref{F3 H3 KT}):
\beq
\Delta C_2\,=\,{nN_f\over 8}\,\,\Big[\,
(\psi-\psi^*)\,(\,\sin\theta_1 d\theta_1\wedge d\varphi_1\,-\,
\sin\theta_2\,d\theta_2\wedge d\varphi_2\,)\,-\,\cos\theta_1\cos\theta_2\,
d\varphi_1\wedge d\varphi_2\,\Big]\,\,,
\label{changeC2}
\eeq
where $\psi^*$ is a constant.  In the study of the R-symmetry anomaly of
the next subsection it will be convenient
to know the change of $C_2$ on the submanifold 
$\theta_1=\theta_2=\theta$,  $\varphi_1=2\pi-\varphi_2=\varphi$. Denoting
by  $C_2^{eff}$  the RR potential $C_2$ restricted to this
cycle, we get from (\ref{changeC2}) that: 
\beq
\Delta C_2^{eff}\,=\,{{n N_f} \over 4} (\psi - \psi^*) \, 
\sin\theta d\theta\wedge d\varphi\,\,. \label{DeltaC2}
\eeq

Let us now study how the Page charges change under these large
gauge transformations. From the expressions written in (\ref{D5D3Page}), we
obtain:
\begin{equation}
\begin{split}
\Delta Q^{Page}_{D5}&=  -{1 \over {4 \pi^2}} \int_{S^3} 
\Delta B_2 \wedge F_1\,\, ,\\
\Delta Q^{Page}_{D3}&={1 \over {(4 \pi^2)^2}} \int_{M_5} 
\Big(-\Delta B_2 \wedge F_3 + \Delta B_2 \wedge B_2 \wedge F_1
+{1\over 2}\,\Delta B_2\wedge\Delta B_2\wedge F_1\,\Big)\,.
\end{split}   \label{DeltaQs}
\end{equation}
By using  in (\ref{DeltaQs}) our ansatz for $F_3$ and $B_2$ \eqref{ansatz}, together
with the expression of $ \Delta B_2 $ given in (\ref{large}) as well as
the relations \eqref{intS3} and \eqref{int5g}, one readily gets:
\begin{equation}
\begin{split}
\Delta Q^{Page}_{D5}  \,&=\, n {N_f \over 2} \,\, ,\\
\Delta Q^{Page}_{D3} \,&=\, n \, M \,+\,n^2 {N_f \over 4}
\, \,. 
\end{split} \label{Delta-Page}
\end{equation}
Thus, under a  large gauge transformation (\ref{large}) with $n=1$, the Page charges
transform as:
\begin{equation}
\begin{split}
Q^{Page}_{D5} &\to Q^{Page}_{D5} \,+\,  {N_f \over 2} \,\, ,\\
Q^{Page}_{D3} &\to Q^{Page}_{D3} \,+\, \, M \,+ {N_f \over 4}
\,\,.
\end{split}\label{Page-change}
\end{equation}
Recall that for our ansatz $Q^{Page}_{D5}=M$ and $Q^{Page}_{D3}=N_0$ (see eqs.
(\ref{QD5=M}) and (\ref{QD3=N0}) ). Thus, eq. (\ref{Page-change}) gives how these constants change under a large gauge transformation. At a given holographic scale $\tau$ we should perform
as many large transformations as needed to have $b_0\in[0,1]$.  Given that $b_0$ grows when 
the holographic coordinate increases, the transformation (\ref{Page-change}) should correspond to the change of ranks under a Seiberg duality when we flow towards the UV. By comparing (\ref{Page-change}) with our previous expressions one can show that this is the case. Actually, one can get an explicit expression of $Q^{Page}_{D5} $ and $Q^{Page}_{D3}$ in terms of the ranks $r_l$ and $r_s$ and the number of flavors $N_{fl}$ and $N_{fs}$. In order to verify this fact, let us suppose that we are in a region of the holographic coordinate such that the two functions $f$ and $k$ of our ansatz are equal. Notice that for the flavored KS solution this happens in the UV, while for the flavored KT this condition 
holds for all values of the radial coordinate. If $f=k$ the normalized flux $b_0$ in (\ref{b_0}) can be written as:
\beq
b_0(\tau)\,=\,{M\over \pi}\,f(\tau)\,\,.
\eeq
Using this expression we can  write the D5-brane Page charge (\ref{QD5-Meff}) as:
\beq
Q^{Page}_{D5}\,=\,M_{eff}\,-\,{N_f\over 2}\,\,b_0\;.
\label{QD5UV}
\eeq
Notice also that the supergravity expression (\ref{Meff}) of $M_{eff}$ can be written  when $f=k$ as:
\beq
M_{eff}\,=\,M\,+\,{N_f\over 2}\,\,b_0\,\,.
\label{Meff-M-UV}
\eeq
Let us next assume that we have chosen our gauge such that, at the given holographic scale, 
$b_0\in [0,1]$.  In that case we can use the value of $M_{eff}$ obtained by the field theory calculation of subsection \ref{charges and ranks} to evaluate the Page charge $Q^{Page}_{D5}$. Actually, by plugging the value of 
 $M_{eff}$ given in (\ref{dictionaryM}) on the right-hand side of (\ref{QD5UV}) we readily get the following
 relation between $Q^{Page}_{D5}$ and the field theory data:
 \beq
 Q^{Page}_{D5}\,=\,r_l\,-\,r_s\,-\,{N_{fl}\over 2}\,\,.
  \label{QD5-ranks}
 \eeq
Similarly, for $f=k$, one can express the D3-brane Page charge (\ref{QD3-Neff}) as:
\beq
 Q^{Page}_{D3}\,=\,N_{eff}\,-\,b_0M\,-\,{b_0^2\over 4}\,N_f\,\,,
 \eeq
which, after using the relation (\ref{Meff-M-UV}), can be written in terms of $M_{eff}$ as:
\beq
 Q^{Page}_{D3}\,=\,N_{eff}\,-\,b_0M_{eff}\,+\,{N_f\over 4}\,b_0^2\,\,.
 \eeq
Again, if we assume that $b_0\in [0,1]$ and use the field theory expressions (\ref{dictionary}) and
 (\ref{dictionaryM})  of $N_{eff}$ and $M_{eff}$, we get:
 \beq
 Q^{Page}_{D3}\,=\,r_s\,+\,{3N_{fl}-N_{fs}\over 16}\,\,.
 \label{QD3-ranks}
 \eeq
Notice that, as it should,  the expressions (\ref{QD5-ranks}) and  (\ref{QD3-ranks})  of 
$Q^{Page}_{D5}$ and $Q^{Page}_{D3}$ that 
we have just found are independent of $b_0$, as far as $b_0\in [0,1]$. Moreover, one can verify that under a field theory Seiberg duality the right-hand sides of 
(\ref{QD5-ranks}) and  (\ref{QD3-ranks}) transform  as the left-hand sides do 
under a large gauge transformation of supergravity.

Finally, let us point out that in this approach Seiberg duality is performed at a fixed 
energy scale and $M_{eff}$ and $N_{eff}$ are left 
invariant (recall that Maxwell charges are gauge invariant). Indeed, by
looking at our ansatz for $B_2$ one easily concludes that the change of
$B_2$ written in (\ref{large}) is equivalent to the following change in
the functions $f$ and $k$
\beq
f\to f-{\pi\over M}\,n\,\,,
\qquad\qquad
k\to k-{\pi\over M}\,n\,\,,
\label{fk-changes}
\eeq
and one can verify that the changes (\ref{Delta-Page}) and
(\ref{fk-changes}) leave the expressions of $M_{eff}$ and $N_{eff}$,
as written in eqs. (\ref{QD5-Meff}) and (\ref{QD3-Neff}),  invariant. 
From eqs.  (\ref{QD5-ranks}) and (\ref{QD3-ranks}) it is clear that the Page charges provide a clean
way to extract the ranks and number of flavors of the corresponding (good)
field theory dual at a given energy scale. Actually, the ranks of this good field
theory description change as  step-like functions along the RG flow, due to the fact that
$b_0$ varies continuosly and needs to suffer a large gauge transformation every time that,
flowing towards the IR,  it reaches the value $b_0=0$ in the good gauge. This large gauge transformation changes $Q^{Page}_{D5}$ and $Q^{Page}_{D3}$ in the way described above, which realizes in supergravity the change of the ranks under a Seiberg duality in field theory.

Let us now focus on a different way of matching the behavior of the field theory and our solutions.

\subsection{R-symmetry anomalies and $\beta$-functions}

We can compute the $\beta$-functions (up to the energy-radius relation) and the R-symmetry anomalies for the two gauge groups both in supergravity and in field theory in the spirit of \cite{Bertolini:2001qa,Klebanov:2002gr,Bertolini:2002xu}. In the UV, where the cascade takes place, they nicely match. For the comparison we make use of the following holographic formulae, which can be derived in the $\mathcal{N}=2$ orbifold case by looking at the Lagrangian of the low energy field theory living on probe (fractional) D3-branes:%
\footnote{We are not completely sure about the sign in the formula relating $C_0$ to the sum of the theta angles. At any rate, with the minus sign we can match the anomaly computations in field theory and in supergravity for the flavored version of Klebanov-Witten's theory proposed in \cite{Ouyang:2003df} (with backreaction of the flavor branes only at first order) and in \cite{Benini:2006hh} (fully backreacting D7-branes). The field theory computation gives $\delta_\varepsilon (\theta_1 +\theta_2)= - N_f\varepsilon $,
which can only be matched with the supergravity results of the two papers using this formula.}
\begin{equation} \begin{split} \label{holographic relations}
\frac{4\pi^2}{g_l^2} + \frac{4\pi^2}{g_s^2} &= \pi\, e^{-\phi} \\
\frac{4\pi^2}{g_l^2} - \frac{4\pi^2}{g_s^2} &= \frac{e^{-\phi}}{2\pi} \Bigl[ \int_{S^2} B_2 - 2\pi^2 \; (\text{mod } 4\pi^2) \Bigr]
\end{split}
\qquad\qquad
\begin{split}
\theta_l^{YM} + \theta_s^{YM} &= - 2\pi \, C_0 \\
\theta_l^{YM} - \theta_s^{YM} &= \frac{1}{\pi} \int_{S^2} C_2 \;.
\end{split}
\end{equation}
 
Strictly speaking, these formulae need to be corrected for small values of the gauge couplings and are only valid in the large 't Hooft coupling regime (see \cite{Benvenuti:2005wi,Strassler:2005qs,Dymarsky:2005xt,Benini:2006hh}), which is the case under consideration. Moreover, they give positive squared couplings only if $b_0=\frac{1}{4\pi^2}\int_{S^2} B_2$ is in the range $[0,1]$. This is the physical content of the cascade: at a given energy scale we must perform a large gauge transformation on $B_2$ in supergravity to shift $\int B_2$ by a multiple of $4\pi^2$ to get a field theory description with positive squared couplings.

We have adapted the indices in \eqref{holographic relations} to the previous convention for the gauge group with the larger (the smaller) rank. Let us restrict our attention to an energy range, between two subsequent Seiberg dualities, where a field theory description in terms of specific ranks holds. In this energy range the gauge coupling $g_l$ of the gauge group with larger rank flows towards strong coupling, while the gauge coupling $g_s$ of the gauge group with smaller rank flows towards weak coupling.
Indeed, as formulae \eqref{holographic relations} confirm, the coupling $g_l$ was not touched by the previous Seiberg duality, starts different from zero and flows to $\infty$ at the end of this range, where a Seiberg duality on its gauge group is needed. The coupling $g_s$ of the gauge group with smaller rank is the one which starts very large (actually 
divergent) after the previous Seiberg duality on its gauge group, and then flows toward weak coupling.

%

In supergravity, due to the presence of magnetic sources for $F_1$, we cannot define a potential $C_0$. Therefore we project our fluxes on the submanifold $\theta_1 = \theta_2\equiv\theta$, $\varphi_1 = 2\pi-\varphi_2\equiv\varphi$, $\forall\, \psi,\tau$ before integrating them. Recalling that $F_3 = dC_2 + B_2\wedge F_1$, what we get from \eqref{F1}-\eqref{ansatz} (in the UV limit) are the effective potentials
\begin{equation} \label{C0 C2 eff}
C_0^{eff} = \frac{N_f}{4\pi} \, (\psi - \psi_0^*) \qquad\qquad \tilde{C}_2^{eff} = \Bigl[\frac{M}{2} + \frac{n N_f}{4}\Bigr] \, (\psi - \psi_2^*) \, \sin{\theta}\,d\theta\wedge d\varphi \;.
\end{equation}
The integer $n$ in $\tilde{C}_2^{eff}$ comes from a large gauge transformation on $B_2$ (Seiberg duality in field theory, see eq. \ref{DeltaC2}) which shifts $b_0 (\tau) \in [n,n+1]$ by $n$ units - so that the gauge transformed $\tilde{b}_0 (\tau)= b_0 (\tau) - n$ is between 0 and 1 - and at the same time shifts $d C_2^{eff}\to d \tilde{C}_2^{eff} = d C_2^{eff} + \pi n \frac{N_f}{4\pi} \sin\theta \,d\theta\wedge d\varphi\wedge d\psi$, since $F_3$ is gauge-invariant, but leaves $C_0$ invariant.

The field theory possesses an anomalous R-symmetry which assigns charge $\frac{1}{2}$ to all chiral superfields.%
\footnote{Although the R-charges of the chiral superfields are half-integer, an R-rotation of parameter $\varepsilon=2\pi$ coincides with a baryonic rotation of parameter $\alpha=\pi$. It follows that $U(1)_R\times U(1)_B$ is parameterized by $\varepsilon\in[0,2\pi]$, $\alpha\in[0,2\pi]$.}
The field theory R-anomalies are easily computed. Continuing to use $r_l$ ($r_s$) for the larger (smaller) group rank and $N_{fl}$ ($N_{fs}$) for the corresponding flavors (see Figure \ref{quiverN}), the anomalies under a $U(1)_R$ rotation of parameter $\varepsilon$ are:
\begin{equation}
\text{Field theory:} \qquad\qquad
\begin{aligned}
\delta_\varepsilon \theta_l &= [2(r_l - r_s) - N_{fl}] \, \varepsilon \\
\delta_\varepsilon \theta_s &= [-2(r_l - r_s) - N_{fs}] \, \varepsilon \;.
\end{aligned} \end{equation}
Along the cascade of Seiberg dualities, the coefficients of the anomalies for the two gauge groups change when we change the effective description; what does not change is the unbroken subgroup of the R-symmetry group.  
Because we want to match them with the supergravity computations, it will be convenient to rewrite the field theory anomalies in the following form:
\begin{equation}
\text{Field theory:} \qquad\qquad
\begin{aligned}
\delta_\varepsilon (\theta_l+\theta_s) &= - N_{f}\, \varepsilon \\
\delta_\varepsilon (\theta_l-\theta_s) &= [4(r_l - r_s) + N_{fs}-N_{fl}] \, \varepsilon \;. \label{anomalies FT}
\end{aligned} \end{equation}

An infinitesimal $U(1)_R$ rotation parameterized by $\varepsilon$ in field theory corresponds to a shift $\psi \to \psi + 2\varepsilon$ in the geometry. Therefore, making use of \eqref{C0 C2 eff}, we find on the supergravity side:
\begin{equation}
\text{SUGRA:} \qquad\qquad
\begin{aligned}
\delta_\varepsilon (\theta_l + \theta_s) &=  - N_f \, \varepsilon \\
\delta_\varepsilon (\theta_l - \theta_s) &= [4M + 2n \, N_f] \, \varepsilon \;.
\end{aligned} \label{anomalies sugra} \end{equation}
These formulae agree with those computed in the field theory.
For the difference of the anomalies, what we can compute and compare is its change after a step in the duality cascade. 
Notice indeed that the difference of the anomalies, as computed in \eqref{anomalies sugra}, gives a step function: as we flow towards the IR, after some energy scale (the scale of a Seiberg duality along the cascade) we need to perform a large gauge transformation in supergravity to turn to the correct Seiberg dual description of the field theory (the only one with positive squared gauge couplings). This corresponds to changing $n \to n-1$ in \eqref{anomalies sugra}, therefore the coefficient of the difference of the R-anomalies decreases by $2N_f$ units. 
This result is reproduced exactly by the field theory computation \eqref{anomalies FT}. In field theory the difference of the anomalies depends on the quantity $4(r_l-r_s) + N_{fs}-N_{fl} $. Keeping the same conventions adopted in  subsection \ref{cascade subsection} and repeating the same reasoning, it's easy to see that after a step of the cascade towards the IR, this quantity decreases exactly by $2N_f$ units.

The dictionary \eqref{holographic relations} allows us also to compute the $\beta$-functions of the two gauge couplings and check further the picture of the duality cascade.

Since we will be concerned in the cascade, we will make use of the flavored Klebanov-Tseytlin solution of Section \ref{KTflavored}, to which the flavored Klebanov-Strassler solution of Section \ref{KSflavored} reduces in the UV limit.

We shall keep in mind that, at a fixed value of the radial coordinate, we want to shift $b_0=\frac{1}{4\pi^2}\int_{S^2} B_2 $ by means of a large gauge transformation in supergravity in such a way that its gauge transformed $\tilde{b}_0=b_0-[b_0]\equiv b_0-n$ 
\footnote{$n$ is a step-like function of the radial coordinate.}
belongs to $[0,1]$: in doing so, we are guaranteed to be using the good description in terms of a field theory with positive squared gauge couplings.

Recall that  
\begin{align}
e^{-\phi}&=\frac{3N_f}{4\pi}(-\rho)\\
b_0&=\frac{2M}{N_f}\bigg(\frac{\Gamma}{\rho}-1\bigg)\,\,,
 \label{b_0 seiberg dualities}
\end{align}
and the dictionary \eqref{holographic relations}, that we rewrite as:
\begin{align}
\frac{8\pi^2}{g_+^2}\equiv \frac{8\pi^2}{g_l^2}+\frac{8\pi^2}{g_s^2}&=2\pi e^{-\phi} \label{g+}\\
\frac{8\pi^2}{g_-^2}\equiv \frac{8\pi^2}{g_l^2}-\frac{8\pi^2}{g_s^2}&=2\pi e^{-\phi}(2\tilde{b}_0-1)\;,\label{g-}
\end{align}
where $\tilde{b}_0\equiv b_0- [b_0] \in [0,1]$ comes from integrating on the two-cycle the suitably gauge transformed Kalb-Ramond potential which must be used in order to describe the correct field theory effective degrees of freedom at the energy scale dual to the value of the radial coordinate.

Then we can compute the following `radial' $\beta$-functions from the gravity dual: 
\begin{align}
\beta^{(\rho)}_+ \equiv \beta^{(\rho)}_{\frac{8\pi^2}{g_+^2}} &\equiv\frac{d}{d\rho}\frac{8\pi^2}{g_+^2}\\
\beta^{(\rho)}_- \equiv \beta^{(\rho)}_{\frac{8\pi^2}{g_-^2}} &\equiv\frac{d}{d\rho}\frac{8\pi^2}{g_-^2}\;,
\end{align}
and we would like to match these with the field theory computations.\\
Using the expressions \eqref{g+}-\eqref{g-}, we can conclude that 
\begin{align}  \label{beta grav1} 
\beta^{(\rho)}_+ &=-3 \frac{N_f}{2} \\   
\beta^{(\rho)}_- &= 3\bigg(\frac{N_f}{2}+Q\bigg) \;,   \label{beta grav2}
\end{align} 
where $Q=N_f [b_0(\rho)]+2M = N_f n(\rho) +2M $ is a quantity which undergoes a change $Q \rightarrow Q-N_f$ as $b_0(\rho)\rightarrow b_0(\rho')=b_0(\rho)-1$ (one Seiberg duality step along the cascade towards the IR), or equivalently $n(\rho)\rightarrow n(\rho')=n(\rho)-1$. 
Up to an overall factor of 2, $Q$ is the same quantity appearing in the difference of the R-anomalies in \eqref{anomalies sugra}.

The field theory computations of the $\beta$-functions give:
\begin{align}
\beta_l &\equiv \beta_{\frac{8\pi^2}{g_l^2}}= 3 r_l-2r_s (1-\gamma_A) - N_{fl} (1-\gamma_q)\\
\beta_s &\equiv \beta_{\frac{8\pi^2}{g_s^2}}= 3 r_s-2r_l (1-\gamma_A) - N_{fs} (1-\gamma_q)\;,
\end{align}
with the usual conventions. Hence
\begin{align}
\beta_+ &\equiv \beta_{l}+\beta_s= (r_l+r_s)(1+2\gamma_A)-N_f(1-\gamma_q)\\
\beta_- &\equiv \beta_{l}-\beta_s= (5-2\gamma_A)(r_l-r_s)+(N_{fs}-N_{fl})(1-\gamma_q)\;.
\end{align}
In order to match the above quantities with the gravity computations \eqref{beta grav1}-\eqref{beta grav2}, an energy-radius relation is required. This is something we miss here.
Although it is not really needed to extract from our supergravity solutions the qualitative information on the running of the gauge couplings, we are going initially to make two assumptions, which can be viewed as an instructive simplification. Let us then assume that the radius-energy relation is $\rho=\ln\frac{\mu}{E_{UV}}$, where $E_{UV}$ is the scale of the UV cutoff dual to the maximal value of the radial coordinate $\rho=0$, and that the anomalous dimensions do not acquire subleading corrections. Matching $\beta_+$ implies $\gamma_A=\gamma_q=-\frac{1}{2}$. Matching $\beta_-$, once we insert these anomalous dimensions, implies that $Q=2(r_l-r_s)-N_{fl}$. 
This quantity correctly shifts as $Q \rightarrow Q-N_f$ when $b_0\rightarrow b_0-1$. 
This last observation allows us to check the consistency of the cascade of Seiberg dualities also against the running of the gauge couplings.%
%

Actually, the qualitative picture of the RG flow in the UV can be extracted from our supergravity solution  even without knowing the precise radius-energy relation, but simply recalling that the radius must be a monotonic function of the energy scale.

It is interesting to notice the following phenomenon: as we flow up in energy and  approach the far UV
$\rho\to 0^-$ in \eqref{b_0 seiberg dualities}, a large number of Seiberg dualities is needed to keep $b_0$ varying in the interval $[0,1]$. The Seiberg dualities pile up the more we approach the UV cut-off $E_{UV}$.
Meanwhile, formula \eqref{beta grav2} reveals that, when going towards the UV cutoff $E_{UV}$, the `slope' in the plots of $\frac{1}{g_i^2}$ versus the energy scale becomes larger and larger, and \eqref{beta grav1} reveals that the sum of the inverse squared gauge coupling goes to zero at this UV cutoff. At the energy scale $E_{UV}$ the effective number of degrees of freedom needed for a weakly coupled description of the gauge theory becomes infinite.
Since $\rho=0$ is at finite proper radial distance from any point placed in the interior $\rho<0$, $E_{UV}$ is a finite energy scale. 

\begin{figure}[ht]
\begin{center}
\includegraphics[width=0.7\textwidth]{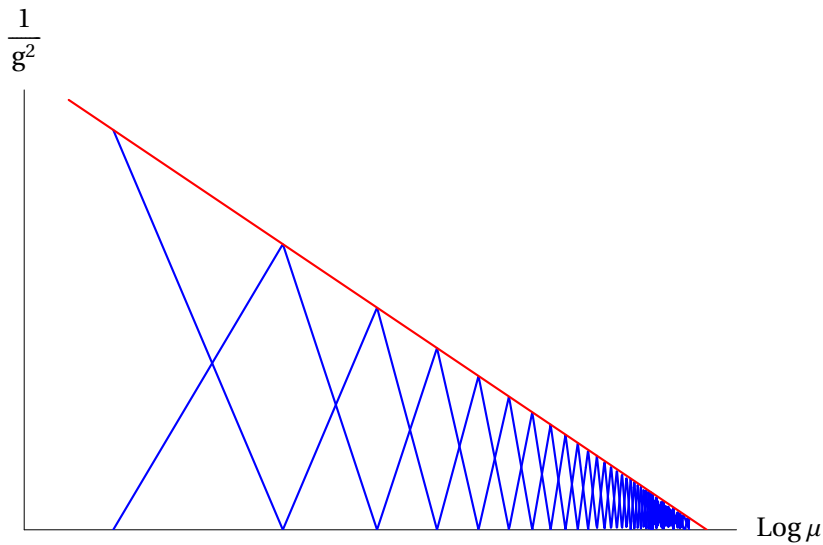}
\end{center}
\caption[wall]{Qualitative plot of the running gauge couplings as functions of the logarithm of the energy scale in our cascading gauge theory. The blue lines are the inverse squared gauge couplings, while the red line is their sum. \label{wall}}
\end{figure}

The picture which stems from our flavored Klebanov-Tseytlin/Strassler solution is that $E_{UV}$ is a so-called ``Duality Wall", namely an accumulation point of energy scales at which a Seiberg duality is required in order to have a weakly coupled description of the gauge theory \cite{Strassler:1996ua}. Above the duality wall, Seiberg duality does not proceed and a weakly coupled dual description of the field theory is not known. See Figure \ref{wall}.

Duality walls were studied in the context of quiver gauge theories first by Fiol \cite{Fiol:2002ah} and later in a series of papers by Hanany and collaborators \cite{Hanany}. Their analysis of this phenomenon was in the framework of quiver gauge theories with only bifundamental chiral superfields, and was restricted to the field theory.

To our knowledge, our solutions are the first explicit realizations of this exotic ultraviolet phenomenon on the supergravity side of the gauge/gravity correspondence.

\section{Final remarks} \label{Sect: final remarks}

In this paper we have presented a very precise example of the duality between field theories with flavors and string solutions that include the dynamics of (flavor) branes. We focused on the Klebanov-Tseytlin/Strassler case, providing a well defined dual field theory, together with different matchings that include the cascade of Seiberg dualities, beta functions and anomalies.
Indeed, we have shown in detail how the ranks of the gauge groups change from a string theory viewpoint (in perfect agreement with the usual field theory prescription), providing also a rigorous definition of the gauge groups ranks in terms of Page charges.
We have also shown how the runnings of the gauge couplings are matched by the string background and how global anomalies are also captured by our solution.

In a future publication \cite{us}, we will present many details concerning the brane embedding and dual field theory. We will also provide more general solutions to our system of BPS equations and analyze the details of their dual dynamics. The interesting IR dynamics (the last steps of the cascade, leading to a baryonic branch of the field theory, behavior of the Wilson loop, etc) will also be spelled out in detail in the forthcoming paper mentioned above.

Many other things can be done with the solution presented here. The study
of implications of these new backgrounds to cosmology and D-brane inflation seems a natural project. On the supergravity side, finding new solutions describing the motion along the baryonic branch of this field theory \cite{Butti:2004pk}, finding and studying the dynamics of the massless Goldstone mode (that should exist) \cite{Gubser:2004qj}; what determines the dual to baryonic operators and their VEV \cite{Benna:2006ib} and of course, the possibility of softly breaking SUSY and studying the new dynamics \cite{Kuperstein:2003yt}, are some of the ideas that naturally come to mind given these new solutions.

\section*{Acknowledgments}
We would like to thank Daniel Are\' an, Matteo Bertolini, Roberto Casero, Paolo Di Vecchia, Jose Edelstein, Roberto Emparan, Jarah Evslin, Amihay Hanany, Igor Klebanov, 
Paolo Merlatti, Sameer Murthy, Angel Paredes, 
Rodolfo Russo, Radu Tatar, Angel Uranga and Alberto Zaffaroni for useful discussions related to this project.
S.C. thanks especially Matteo Bertolini for discussions, comments and feedback on the draft.
F.B., S.C. and C.N. are grateful to the Galileo Galilei Institute for Theoretical Physics in Florence for the hospitality and the INFN for partial support during the completion of this work.
The  work of FC and AVR was
supported in part by MCyT and  FEDER  under grant
FPA2005-00188,  by Xunta de Galicia (Conselleria de Educacion and grant PGIDIT06PXIB206185PR)
and by  the EC Commission under  grant MRTN-CT-2004-005104. F.B. and S.C are partially supported by Italian MIUR under contract PRIN-2005023102.



\begin{thebibliography}{99}



\bibitem{'tHooft:1973jz}
  G.~'t Hooft,
  Nucl.\ Phys.\  B {\bf 72}, 461 (1974).

\bibitem{Maldacena:1997re}
  J.~M.~Maldacena,
  Adv.\ Theor.\ Math.\ Phys.\  {\bf 2}, 231 (1998)
  [Int.\ J.\ Theor.\ Phys.\  {\bf 38}, 1113 (1999)]
  [arXiv:hep-th/9711200].
\bibitem{Itzhaki:1998dd}
  N.~Itzhaki, J.~M.~Maldacena, J.~Sonnenschein and S.~Yankielowicz,
  Phys.\ Rev.\  D {\bf 58}, 046004 (1998)
  [arXiv:hep-th/9802042].
\bibitem{Gubser:1998bc}
  S.~S.~Gubser, I.~R.~Klebanov and A.~M.~Polyakov,
  Phys.\ Lett.\  B {\bf 428}, 105 (1998)
  [arXiv:hep-th/9802109].
\bibitem{Witten:1998qj}
  E.~Witten,
  Adv.\ Theor.\ Math.\ Phys.\  {\bf 2}, 253 (1998)
  [arXiv:hep-th/9802150].

  
  
  
\bibitem{Klebanov:1998hh}
  I.~R.~Klebanov and E.~Witten,
  Nucl.\ Phys.\  B {\bf 536}, 199 (1998)
  [arXiv:hep-th/9807080].
\bibitem{Gubser:1998fp}
  S.~S.~Gubser and I.~R.~Klebanov,
  Phys.\ Rev.\  D {\bf 58}, 125025 (1998)
  [arXiv:hep-th/9808075].
\bibitem{Klebanov:1999rd}
  I.~R.~Klebanov and N.~A.~Nekrasov,
  Nucl.\ Phys.\  B {\bf 574}, 263 (2000)
  [arXiv:hep-th/9911096].
\bibitem{Klebanov:2000nc}
  I.~R.~Klebanov and A.~A.~Tseytlin,
  Nucl.\ Phys.\  B {\bf 578}, 123 (2000)
  [arXiv:hep-th/0002159].
\bibitem{Klebanov:2000hb}
  I.~R.~Klebanov and M.~J.~Strassler,
  JHEP {\bf 0008}, 052 (2000)
  [arXiv:hep-th/0007191].
\bibitem{Gubser:2004qj}
  S.~S.~Gubser, C.~P.~Herzog and I.~R.~Klebanov,
  JHEP {\bf 0409}, 036 (2004)
  [arXiv:hep-th/0405282].
\bibitem{Dymarsky:2005xt}
  A.~Dymarsky, I.~R.~Klebanov and N.~Seiberg,
  JHEP {\bf 0601}, 155 (2006)
  [arXiv:hep-th/0511254].
\bibitem{Benna:2006ib}
  M.~K.~Benna, A.~Dymarsky and I.~R.~Klebanov,
  arXiv:hep-th/0612136.
\bibitem{Butti:2004pk}
  A.~Butti, M.~Grana, R.~Minasian, M.~Petrini and A.~Zaffaroni,
  JHEP {\bf 0503}, 069 (2005)
  [arXiv:hep-th/0412187].
\bibitem{Strassler:2005qs}
  M.~J.~Strassler,
  arXiv:hep-th/0505153.



\bibitem{Veneziano:1976wm}
  G.~Veneziano,
  Nucl.\ Phys.\  B {\bf 117}, 519 (1976).







\bibitem{Karch:2002sh}
  A.~Karch and E.~Katz,
  JHEP {\bf 0206}, 043 (2002)
  [arXiv:hep-th/0205236].

\bibitem{Ouyang:2003df}
  P.~Ouyang,
  Nucl.\ Phys.\  B {\bf 699}, 207 (2004)
  [arXiv:hep-th/0311084].

\bibitem{Kuperstein:2004hy}
  S.~Kuperstein,
  JHEP {\bf 0503}, 014 (2005)
  [arXiv:hep-th/0411097].
  


\bibitem{Casero:2006pt}
  R.~Casero, C.~Nunez and A.~Paredes,
  Phys.\ Rev.\  D {\bf 73}, 086005 (2006)
  [arXiv:hep-th/0602027].
\bibitem{Paredes:2006wb}
  A.~Paredes,
  JHEP {\bf 0612}, 032 (2006)
  [arXiv:hep-th/0610270].
\bibitem{Casero:2007pz}
  R.~Casero and A.~Paredes,
  arXiv:hep-th/0701059.


\bibitem{Benini:2006hh}
  F.~Benini, F.~Canoura, S.~Cremonesi, C.~Nunez and A.~V.~Ramallo,
  JHEP {\bf 0702}, 090 (2007)
  [arXiv:hep-th/0612118].

\bibitem{us}
F.~Benini, F.~Canoura, S.~Cremonesi, C.~Nunez and A.~V.~Ramallo, in preparation.


\bibitem{MarolfCB}
  D.~Marolf,
  arXiv:hep-th/0006117.



\bibitem{Page}
D. D. Page, Phys. Rev. {\bf 28}, 2796 (1983).


\bibitem{Strassler:1996ua}
  M.~J.~Strassler,
{\it Prepared for International Workshop on Perspectives of Strong Coupling Gauge Theories (SCGT 96), Nagoya, Japan, 13-16 Nov 1996}





\bibitem{Radu}
  F.~Benini, S.~Cremonesi, R.~Tatar, in preparation.



\bibitem{Seiberg:1994pq}
  N.~Seiberg,
  Nucl.\ Phys.\  B {\bf 435}, 129 (1995)
  [arXiv:hep-th/9411149].
  
\bibitem{Franco:2006es}
  S.~Franco and A.~M.~Uranga,
  JHEP {\bf 0606}, 031 (2006)
  [arXiv:hep-th/0604136].

\bibitem{Bertolini:2000dk}
  M.~Bertolini, P.~Di Vecchia, M.~Frau, A.~Lerda, R.~Marotta and I.~Pesando,
  JHEP {\bf 0102}, 014 (2001)
  [arXiv:hep-th/0011077].

\bibitem{Bertolini:2001qa}
  M.~Bertolini, P.~Di Vecchia, M.~Frau, A.~Lerda and R.~Marotta,
  Nucl.\ Phys.\  B {\bf 621}, 157 (2002)
  [arXiv:hep-th/0107057].

\bibitem{Aspinwall:1995zi}
  P.~S.~Aspinwall,
  Phys.\ Lett.\  B {\bf 357}, 329 (1995)
  [arXiv:hep-th/9507012].
  

\bibitem{Morrison:1998cs}
  D.~R.~Morrison and M.~R.~Plesser,
  Adv.\ Theor.\ Math.\ Phys.\  {\bf 3}, 1 (1999)
  [arXiv:hep-th/9810201].
  

\bibitem{Grana:2001xn}
  M.~Grana and J.~Polchinski,
  Phys.\ Rev.\  D {\bf 65}, 126005 (2002)
  [arXiv:hep-th/0106014].
  
\bibitem{Polchinski:2000mx}
  J.~Polchinski,
  Int.\ J.\ Mod.\ Phys.\  A {\bf 16}, 707 (2001)
  [arXiv:hep-th/0011193].

\bibitem{Bershadsky:1995qy}
  M.~Bershadsky, C.~Vafa and V.~Sadov,
  Nucl.\ Phys.\  B {\bf 463}, 420 (1996)
  [arXiv:hep-th/9511222].
 
\bibitem{Klebanov:2002gr}
  I.~R.~Klebanov, P.~Ouyang and E.~Witten,
  Phys.\ Rev.\  D {\bf 65}, 105007 (2002)
  [arXiv:hep-th/0202056].
  
\bibitem{Bertolini:2002xu}
  M.~Bertolini, P.~Di Vecchia, M.~Frau, A.~Lerda and R.~Marotta,
  Phys.\ Lett.\  B {\bf 540}, 104 (2002)
  [arXiv:hep-th/0202195].
  
    
  
\bibitem{Benvenuti:2005wi}
  S.~Benvenuti and A.~Hanany,
  JHEP {\bf 0508}, 024 (2005)
  [arXiv:hep-th/0502043].
  

  

  
 
\bibitem{Fiol:2002ah}
  B.~Fiol,
  JHEP {\bf 0207}, 058 (2002)
  [arXiv:hep-th/0205155].
  
    
  
  
  
  
  \bibitem{Hanany}
  A.~Hanany and J.~Walcher,
  JHEP {\bf 0306}, 055 (2003)
  [arXiv:hep-th/0301231].
 S.~Franco, A.~Hanany, Y.~H.~He and P.~Kazakopoulos,
  [arXiv:hep-th/0306092].
  
  
  
  
      

\bibitem{Kuperstein:2003yt}
  S.~Kuperstein and J.~Sonnenschein,
  JHEP {\bf 0402}, 015 (2004)
  [arXiv:hep-th/0309011].
  M.~Schvellinger,
  JHEP {\bf 0409}, 057 (2004)
  [arXiv:hep-th/0407152].
  
    
    
    
    
    
    
    
    
    
    
    
  
\end{thebibliography}
\end{document}